\documentclass[a4paper,12pt,oneside]{article}
\usepackage[latin1]{inputenc}
\usepackage{amsmath}
\usepackage{amsfonts}
\usepackage{amssymb}
\usepackage{graphicx}
\usepackage{footnote}
\usepackage{float}
\usepackage{subcaption}
\DeclareGraphicsExtensions{.bmp,.png,.pdf,.jpg,.eps}
\usepackage{physics}
\usepackage{mathrsfs}

\usepackage[colorlinks]{hyperref}
\hypersetup{
    colorlinks=true,
    linkcolor=blue,
    filecolor=magenta,  
    urlcolor=[rgb]{0.00,0.00,0.50},    
    citecolor=red,
    }

\usepackage{url}

\advance \textheight by \topskip
\usepackage{amsmath}
\usepackage[english]{babel}
\makeatletter
 \@addtoreset{equation}{section}
\makeatother
\usepackage[latin1]{inputenc}
\usepackage{amsmath}
\numberwithin{equation}{section}

\advance \textheight by \topskip
\usepackage{amsmath}
\usepackage[english]{babel}
\DeclareMathAlphabet\mathbfcal{OMS}{cmsy}{b}{n}
\DeclareMathAlphabet{\boldmathe}{T1}{cmr}{bx}{it}
\newcommand{\mbf}[1]{\boldmathe{#1}}
\newcommand{\mbfgr}[1]{\textit{\mbox{\boldmath$#1$}}}

\def\vA{\mbf{A}}
\def\vB{\mbf{B}}

\def\vS{\mbf{S}}

\def\vx{\mbf{x}}

\def\vp{\mbf{p}}

\def\vmu{\mbfgr{\mu}}

\def\be{\begin{equation}}
\def\ee{\end{equation}}

\def\R{\mathbb R}

\def\H{\mathbb H}
\def\I{\mathbb{I}}

\def\be{\begin{equation}}
\def\ee{\end{equation}}

\def\R{\mathbb R}

\def\H{\mathbb H}
\def\I{\mathbb{I}}

\def\vA{\mbf{A}}
\def\vB{\mbf{B}}

\def\vp{\mbf{p}}

\def\be{\begin{equation}}
\def\ee{\end{equation}}

\def\Z{\mathbb Z}

\def\R{\mathbb R}

\def\H{\mathbb H}
\def\I{\mathbb{I}}

\usepackage{todonotes}

\begin{document}
%%%%%%%%%%%%%%%%%%%%%%%%%%%%%

\title{
{\bf Conformal bridge transformation, \\ $\mathcal{PT}$- and super- symmetry}}

\author{{\bf  Luis Inzunza and Mikhail S. Plyushchay} 
 \\
[8pt]
{\small \textit{Departamento de F\'{\i}sica,
Universidad de Santiago de Chile, Av. V\'ictor Jara 3493, Santiago,
Chile  }}\\
[4pt]
 \sl{\small{E-mails:   
\textcolor{blue}{luis.inzunza@usach.cl},
\textcolor{blue}{mikhail.plyushchay@usach.cl}
}}
}
\date{}
\maketitle

\begin{abstract}
 Supersymmetric extensions of the 1D and 2D  Swanson models are investigated by applying  the conformal 
 bridge transformation (CBT) to the  first order Berry-Keating Hamiltonian multiplied by $i$ 
 and its conformally neutral  enlargements. 
 The CBT plays the role of the Dyson map that transforms the models into supersymmetric 
generalizations of the 1D and 2D harmonic oscillator systems, allowing us to define pseudo-Hermitian 
conjugation and a suitable inner product.   In the 1D case, we  construct a $\mathcal{PT}$-invariant 
supersymmetric model with $N$  subsystems  by using the conformal generators 
of supersymmetric free particle,  and identify its complete set of 
the true bosonic and fermionic integrals of motion.  We also investigate an exotic 
 $N=2$ supersymmetric generalization, in which the higher order supercharges generate 
nonlinear superalgebras.  We generalize the construction for the 2D case to obtain the $\mathcal{PT}$-invariant     
supersymmetric  systems that transform into the spin-1/2 Landau problem with and without an 
additional Aharonov-Bohm flux,  where in the  latter case, the well-defined integrals of motion 
appear only when the flux is quantized.  We also build a 2D supersymmetric Hamiltonian 
related to the ``exotic rotational invariant harmonic oscillator"  system governed 
 by a dynamical parameter $\gamma$. The   bosonic and fermionic  hidden symmetries for this model
 are shown to exist  for rational values of $\gamma$.

\vskip.5cm\noindent
\end{abstract}

\section{Introduction}
The focus on non-Hermitian $\mathcal{PT}$-invariant systems 
has grown significantly following the pioneering studies  of Bender and collaborators 
\cite{Bender1998,Bender1998(2),Bender2002}. 
Such systems 
attract attention and find applications 
 in various areas of physics, including
optics 
\cite{PTOptics1,PTOptics2,PTOptics3,PTOptics4,PTconfDar}, 
condensed matter physics \cite{PTCDM3,PTCDM1,PTCDM2}, 
quantum field theory  
\cite{PTQFT,BenderQFT1,BenderQFT2,PTQCD,PTSB1,PTSB2,AlElMil,PTBPS1},  
gravity and cosmology \cite{Mannh,Mannh2},
nonlinear waves \cite{KoYaZ},
and theory of integrable models 
of finite \cite{PTdef1,PTdef2,PTdef3,PTdef3(2),PTdef3(3)} 
and infinite number of degrees of freedom  
\cite{NLSWE,NLSWE(2),ABsystem1,ABsystem1(2),ABsystem3,Cen,Cen(2),Hirota2,CenFring2016,Correa2016,Cen2016,MatMik1,Cen2017}. 

The interest in them
 is based on the presence of 
non-Hermitian terms in their Hamiltonians, 
which, in particular, 
can be interpreted as the result of interaction of the respective
systems
with the environment \cite{Bender1998,SGH,PT2,PT3,PT3(2),PT4}. 
When such a  Hamiltonian is pseudo-Hermitian (see Ref. \cite{PT2} and Sec. \ref{SecSwanson} below), 
it can be related to a Hermitian system, that guarantees a unitarity of  evolution.

A good starting point for studying $\mathcal{PT}$-invariant  systems is the Swanson oscillator
\cite{Swanson2004,MorFri2006,Sm2006,Sm2008,Ali}, a one-dimensional 
multi-parameter toy model. This system has 
all real eigenvalues for a certain 
 choice  of  the  parameters. 
Consequently, one can construct the Dyson map \cite{Dyson} that transforms the corresponding 
non-Hermitian Hamiltonian into a 
Hermitian one 
in such a case.  

The present article aims to construct 
supersymmetric extensions 
of this model 
 on the real
line and of its generalization in the Euclidean plane. 
The resulting  systems 
are  interesting for
 their connection with the spin-1/2 Landau  problem, 
the models in the presence of Aharonov-Bohm 
 flux associated with 
 anyons \cite{
LeiMyr,MacWil,Hall,CalAny1}, 
 the physics of Bose-Einstein condensates in
rotating harmonic traps \cite{BoseHarm1,ERIHO},
and gravitoelectromagnetism \cite{Inzunza:2021vlb}.

The usual construction  
of one-dimensional supersymmetric  quantum models
 is  based on 
 the Darboux \cite{MatSal,Cooper}
and confluent Darboux transformations \cite{PTconfDar,Inzunza:2019xml}, 
and in Refs. \cite{SUSYSwanson1,SUSYSwanson2} 
some interesting results related 
to super-extended generalized Swanson systems  are presented.  
We will take a different approach here   
grounded on  the peculiar property 
 of the Swanson 
model admitting  a representation
 in terms of the 
 conformal $\mathfrak{so}(2,1)$  generators of the free particle system, 
 that allows us 
  to use the recently developed conformal bridge transformation (CBT) technique
 \cite{Inzunza:2019sct}.
This mapping  relates the symmetry generators  
 and eigenstates of asymptotically  free and harmonically confined versions
 of the $\mathfrak{so}(2,1)\cong\mathfrak{sl}(2,\R)$ 
 conformal mechanics models (e.g., of the free particle and the 
 harmonic oscillator in the simplest case).
In this way, we identified hidden symmetries and 
spectra of different harmonically trapped systems  in a monopole \cite{Inzunza:2020ogw} 
and cosmic string backgrounds \cite{InzPly7,Inzunza:2021vlb}, and constructed 
exotic rotational invariant  harmonic oscillator (ERIHO)
related to the physics of Bose-Einstein condensates \cite{ERIHO}
and gravitoelectromagnetism \cite{Inzunza:2021vlb}.
The CBT was also  applied  
to the study of the black holes mechanics in ref.  \cite{AchLiv}.

As a basis 
for the aim of the present research, we will go deep into
%use   
the link between 
the CBT and the $\mathcal{PT}$-symmetric Swanson oscillator  observed 
%established  
by us recently  in \cite{InzPly9}.

The 
article is organized as follows.
 In Sec.  \ref{SecSwanson} we briefly review the one-dimensional Swanson model and its connection 
 with the CBT, which serves as the Dyson map. 
 We also provide there some comments on the direct higher dimensional generalization of the model. 
  The $N$-supersymmetric extension 
 and a higher order $N=2$ supersymmetric extension of the one-dimensional Swanson model
 are discussed in Sec.  \ref{SecSUSYCBT1d}, 
 where  the first and higher order bosonic and  fermionic 
 integrals of motion for these systems are 
  identified, and the corresponding superalgebraic structures are discussed.  
Sec.  \ref{SecSUSYCBT2d} is devoted to generalization 
    of the construction to
the two-dimensional case. First, we build 
a two-dimensional 
$\mathcal{PT}$-symmetric model using the generators of the planar
supersymmetric free particle system  obtained
via a non-relativistic limit from the free Dirac and Klein-Gordon equations. 
The resulting model is related to the spin-1/2 Landau problem 
by applying 
the CBT.   
After that, we study two 
generalizations of this non-Hermitian  $\mathcal{PT}$-invariant 
supersymmetric model. 
The first   generalization is constructed by  
adding an Aharonov-Bohm flux,  for which
integrals of motion
are obtained from the free particle system  
only when the flux is quantized. 
Another
 generalized system   is related by the CBT   
 to the supersymmetric extension of the
ERIHO system. 
In concluding Sec.  \ref{SecDisc} 
we discuss 
some related interesting 
open problems.

%%%%%%%%%%%%%%%%%%%%%%
%%%%%%%%%%%%%%%%%%%%%%%%%%%%%%%%%%%%%%%%%%%%%%%%%%%%

\section{Swanson model and CBT}
\label{SecSwanson}

The non-Hermitian Swanson model is given by 
 the $\mathcal{PT}$-symmetric 
Hamiltonian operator \cite{Swanson2004}
\begin{eqnarray}
&\label{SwansonH}
\hat{H}_{\alpha,\beta,\omega}=\hbar  \left[\omega(\hat{a}_\omega^{+}\hat{a}_\omega^{-}+\frac{1}{2})+
\alpha(\hat{a}_\omega^-)^{2}+
\beta(\hat{a}_\omega^{+})^{2}\right]\,,&
\end{eqnarray}
 where  $\omega>0$,  $\alpha$ and $\beta$ are real  parameters 
 with the dimension of frequency, 
  subject to
   the relations  $\omega^2-4\alpha\beta>0$, $\alpha\not=\beta$, and
  $ \hat{a}_\omega^\pm=\sqrt{\frac{m\omega}{2\hbar}}(\hat{x}\mp\frac{\hbar}{m\omega}\frac{d}{dx})$
  are the  bosonic harmonic creation and annihilation  
  operators,  $[\hat{a}_\omega^-,\hat{a}_\omega^+]=1$. 
  By a generalized Bogolyubov transformation,
 operator (\ref{SwansonH}) can be transformed into Hamiltonian of the 
  harmonic oscillator with eigenvalues $\hbar\Omega(n+\frac{1}{2})$, 
 $\Omega=\sqrt{\omega^2-4\alpha\beta}$, $n=0,1,\ldots$
 \footnote{In the original article   \cite{Swanson2004} the parameters $\omega$, $\alpha$ and $\beta$ 
 were restricted by the inequality $\omega^2-4\alpha\beta\geq0$.  
 We take a strict inequality in order to exclude the case $\Omega=0$.}.
 Some of such transformations are explicitly considered  in Appendix \ref{AppLabel}. 
This model gained popularity over the years
as
  it allows to 
lay a good base for the study of $\mathcal{PT}$-symmetric 
systems in general,  see  for example
 Refs.
\cite{MorFri2006,Sm2006,Sm2008,Ali}.  
 
 In terms of the Hamiltonian and dynamic generators of  the  
 $\mathfrak{so}(2,1)$
 conformal symmetry~\footnote{
As \emph{dynamic  symmetry  generators}  we mean the integrals
of motion that explicitly depend on time, 
$\frac{d\hat{A}}{dt}=\frac{i}{\hbar}[\hat{H},\hat{A}]+\frac{\partial \hat{A}}{\partial t}=0$.   
  They do not commute with the corresponding Hamiltonian operator,
  and throughout  the
 article such generators are taken
 at $ t = 0 $. 
 Their explicit dependence on time can be 
 restored  by  applying a unitary transformation with the 
 evolution operator. In contrast, the \emph{true integrals of motion}  
 are those
 operators that satisfy the equation
$\frac{d\hat{B}}{dt}=\frac{i}{\hbar}[\hat{H},\hat{B}]=0$. } 
 of the one-dimensional 
free particle system, 
\begin{eqnarray}
&\label{so21gen}
\hat{H}=\frac{1}{2m}\hat{p}^2\,,\qquad 
\hat{D}=\frac{1}{4}\{\hat{x},\hat{p}\}\,,
\qquad 
\hat{K}=\frac{m}{2}\hat{x}^{2}\,,
&\\&
[\hat{D},\hat{H}]=i\hbar\hat{H}\,,\qquad 
[\hat{D},\hat{K}]=-i\hbar\hat{K}\,,\qquad 
[\hat{K},\hat{H}]=2i\hbar\hat{D}\,, \label{so21gen+}
&
\end{eqnarray} 
operator 
$\hat{H}_{\alpha,\beta,\omega} $
reads as 
\begin{eqnarray}
\label{SwansonInSO1}
\hat{H}_{\alpha,\beta,\omega} 
=(1-\omega^{-1}(\alpha+\beta))\hat{H}+(1+\omega^{-1}(\alpha+\beta))\omega^2\hat{K}
+2i(\alpha-\beta)\hat{D}\,. 
\end{eqnarray}
A system of this kind 
is related to the alternative non-Hermitian model  
\begin{eqnarray}
&
\label{SwansonInSO2}
\hat{H}_{\Omega}=\hat{H}+2i\Omega\hat{D}=-\frac{\hbar^2}{2m}\frac{d^2}{dx^2}+\frac{\Omega\hbar }{2}\left(x\frac{d}{dx}+\frac{1}{2}\right)\,,
\end{eqnarray}
via a
similarity transformation. This is 
 easy to see since 
the eigenvalue equation corresponding to operator 
(\ref{SwansonInSO2}) is nothing else than the
 Hermite 
  equation.
Then, 
the similarity transformation $e^{-\frac{\Omega}{\hbar}\hat{K}}\hat{H}_\Omega e^{\frac{\Omega}{\hbar}\hat{K}}$ 
gives us the harmonic oscillator system with frequency $\Omega$,
and  $e^{-\frac{\Omega}{\hbar}\hat{K}}$ 
to be 
 the Gaussian 
factor appearing in the harmonic oscillator wave functions. 
In the usual interpretation 
of $\mathcal{PT}$-symmetric systems, 
$e^{-\frac{\Omega}{\hbar}\hat{K}}$ 
generates the Dyson map and 
its square corresponds to the metric operator, which
in this case is 
the weight
function for  Hermite
polynomials inner product. 
Clearly, a composition of this last mapping  and of  the
generalized  Bogolyubov  transformation mentioned above
relates (\ref{SwansonInSO1}) and (\ref{SwansonInSO2}). 

Now,  the system
(\ref{SwansonInSO2}) can be obtained 
from (\ref{SwansonH})
 in two alternative ways.
First,  this can be done by putting 
$\beta=-\frac{\Omega}{2}$,
$\alpha=\frac{\Omega}{2}+\varepsilon^2$,
$\omega=\varepsilon$,  
and taking the limit $\varepsilon\rightarrow 0$
in
(\ref{SwansonInSO1}).
This limit  produces
divergent operators 
in some constructions of Dyson maps for Swanson model
considered in  \cite{Sm2006}, 
and for that reason 
the system 
(\ref{SwansonInSO2})
is out of sight 
from the common point of 
view~\footnote{
In \cite{Sm2006}, the authors choose to work 
 from the beginning
 with  a Dyson operator of the form 
$S=\exp(\epsilon \hat{a}_\omega^+\hat{a}_\omega^-+\eta(\hat{a}_\omega^-)^{2}+\eta^*(\hat{a}_\omega^-)^{2})$
with $\epsilon^{2}-4|\eta|^2>0$,  
that is incompatible 
with $e^{-\frac{\Omega}{\hbar}\hat{K}}$. 
 The  operator $S$ 
is rewritten  then  in terms of a new parameter $z=z(\alpha,\beta,\omega,\epsilon,\eta)$ 
 that should  take values in
 the interval $(-1,1)$.  
For our  limiting procedure, 
their parameter $z$ goes to infinity, 
see equations (9), (12) and (15) in \cite{Sm2006}
for  the details.}.

Another
way is via the similarity transform
\begin{eqnarray}
&
\label{Swanson2iD}
e^{\frac{\hat{H}}{2\Omega\hbar}} (\hat{H}_{\mathcal{PT}}) e^{-\frac{\hat{H}}{2\Omega\hbar}}=\hat{H}_{\Omega}\,,
\qquad
\hat{H}_{\mathcal{PT}}:=2i\Omega\hat{D}=\frac{\Omega}{2}(x\frac{d}{dx}+\frac{1}{2})\,,
&
\end{eqnarray}
applied to 
the $\mathcal{PT}$-invariant
   first order differential  operator $\hat{H}_{\mathcal{PT}}$, 
   that
   corresponds 
   to the  well known Weierstrass transformation, 
   see \cite{Inzunza:2019sct} and references therein.   
However, the operator $\hat{H}_{\mathcal{PT}}$
cannot be  
 obtained
  from (\ref{SwansonInSO1}) via 
 the described limiting procedure applied to the parameters $\alpha$, $\beta$ and $\omega$.
This can be done, nevertheless,   
if to generalize further the Swanson model by introducing 
 a real parameter $\gamma$ in front 
of the first term in (\ref{SwansonH}) 
and taking the  limit $\gamma\rightarrow 0$,
see \cite{InzPly9}.
The $\mathcal{PT}$-symmetric operator $\hat{H}_{\mathcal{PT}}$
can be
transformed 
 into the harmonic oscillator Hamiltonian 
 via the CBT
\cite{Inzunza:2019sct}~ 
 \begin{eqnarray}
& \label{ConformalBridge1}
 \hat{\mathfrak{S}}_\Omega:= e^{-\frac{\Omega}{\hbar}\hat{K}}
 e^{\frac{\hat{H}}{2\Omega}}e^{\frac{i}{\hbar}\ln(2)\hat{D}}=\exp(\frac{\pi}{4\Omega\hbar}(\hat{H}-\Omega^2\hat{K}))=
 \exp(-\frac{\pi}{8}((\hat{a}_\Omega^-)^2+(\hat{a}_\Omega^+)^2))
 \,,\\
  &
  \hat{\mathfrak{S}}_\Omega (\hat{H}_\mathcal{PT})\hat{\mathfrak{S}}^{-1}_\Omega =\hat{H}+\Omega^2 \hat{K}\,.&
\end{eqnarray} 
Note that the two exponential operators in the first equality are just 
the composition of the explicit similarity transformations mentioned above, 
while 
the additional  
factor $e^{\frac{i}{\hbar}\ln(2)\hat{D}}$ does not change the result and 
 helps to rewrite the 
 CBT generator $\hat{\mathfrak{S}}_\Omega$   in 
the 
 explicitly 
 Hermitian
 form, which looks like the 
 evolution operator of the  inverted 
 harmonic
 oscillator system for complex time 
$t=i\pi/4\Omega $. 
From this line of reasoning, it follows that models (\ref{SwansonH}), (\ref{SwansonInSO2}),
and 
$\hat{H}_{\mathcal{PT}}$ given by Eq. (\ref{Swanson2iD}) 
are similarity equivalent,  and which 
one
of them we work with depends 
on the problem. This 
statement 
is rigorously  re-enforced in Appendix \ref{AppLabel}.

From now on, we will continue 
to work 
with (\ref{Swanson2iD}) 
since 
this operator 
can easily be generalized 
to higher dimensions, as we will see below. 
For simplicity, we also change the notation   
\begin{eqnarray}
\Omega\rightarrow \omega\,,\qquad
\hat{a}^\pm_\Omega\rightarrow \hat{a}^\pm \,,\qquad
\hat{\mathfrak{S}}_\Omega\rightarrow
\hat{\mathfrak{S}} \,.
\end{eqnarray}

The advantage to take the operator (\ref{Swanson2iD}) as a 
 $\mathcal{PT}$-symmetric Hamiltonian
  is  that its
  formal, single-valued but not square integrable on $\R$
 eigenstates
\begin{eqnarray}
\label{statesofsiD}
&
\phi_n(x)=
\frac{1}{\sqrt{2^n n!}}\left( \frac{m\omega}{2\hbar}\right)^{\frac{1}{4}}\left(
\sqrt{\frac{m\omega}{\hbar}}x
\right)^{n}\,,\quad 
 \hat{H}_{\mathcal{PT}}\phi_n(x)=\hbar\omega(n+\frac{1}{2})\phi_n(x)\,, 
  \quad
n=0,1,\ldots\,,\qquad
&
\end{eqnarray}
have a
 simple  monomial  form. 
Furthermore,  these functions  are  Jordan states of the free particle Hamiltonian as well \cite{CarPly,Inzunza:2019xml}
\footnote{The Jordan states of  $\hat{H}$ are those functions that are annihilated by a certain polynomial of 
$\hat{H}$. In this case 
$(\hat{H})^{k+1}\phi_{2k}=0$ and $(\hat{H})^{k+1}\phi_{2k+1}=0$, with $k=0,1,\ldots$.}, 
that
 helps to compute 
their transformation under the action of $\hat{\mathfrak{S}}$, 
\begin{eqnarray}
&\label{harmonicStates}
\hat{\mathfrak{S}}\phi_n(x) 
 =
 \frac{1}{\sqrt{2^n n!}}\big(\frac{m\omega}{\hbar \pi}\big)^{\frac{1}{4}}
  H_{n}\big(\sqrt{\frac{m\omega}{\hbar}}x\big)e^{-\frac{m\omega x^2}{2\hbar}}
 := 
 \psi_{n}(x)\,.&
\end{eqnarray}
Here, $H_{n}(\eta)$ are the Hermite polynomials, and
  the
 resulting functions are the normalized eigenstates of the harmonic oscillator system.

On the other hand, it is easy to see that the operators $\hat{H}$ and $\hat{p}$ 
($\hat{K}$ and $\hat{x}$) are the second and first order lowering (raising) ladder
operators for
formal eigenstates (\ref{statesofsiD}) 
of the $\mathcal{PT}$-invariant 
Hamiltonian
(\ref{Swanson2iD}).
Under the CBT
 generated by   
(\ref{ConformalBridge1}), 
they transform as 
\begin{eqnarray}
&\label{CBT1}
\hat{\mathfrak{S}}
\left(\frac{i}{\sqrt{m\omega\hbar}}\hat{p},
\sqrt{\frac{m\omega }{\hbar}}\hat{x},-\frac{1}{\omega\hbar} \hat{H},\frac{\omega}{\hbar}\hat{K},
\hbar^{-1} i \hat{D}
\right)
\hat{\mathfrak{S}}^{-1}
=
(\hat{a}^{-},\hat{a}^{+},\hat{J}_-,\hat{J}_+,\hat{J}_0)\,,
&
\end{eqnarray}
where
\begin{eqnarray}\label{JJJ}
&
\hat{J}_0=\frac{1}{2}(\hat{a}^+\hat{a}^-+\frac{1}{2})=\frac{1}{2\omega\hbar}\hat{H}_{\text{osc}}\,,
\qquad
\hat{J}_{\pm}=\hat{J}_{1}\pm i\hat{J}_{2}=\frac{1}{2}(\hat{a}^\pm)^2
&
\end{eqnarray}
are the dimensionless generators of the
  $\mathfrak{sl}(2,\R)$ conformal algebra 
\begin{eqnarray}
[\hat{J}_0,\hat{J}_\pm]=\pm \hat{J}_\pm\,,\qquad
[\hat{J}_-,\hat{J}_+]=2\hat{J}_0\,. 
\end{eqnarray}
In effect,
the transformation (\ref{CBT1}) is nothing else than the 
Dyson map \cite{Dyson} that
relates the non-Hermitian Hamiltonian 
(\ref{Swanson2iD}) with  
the harmonic oscillator Hamiltonian $\hat{H}_{\text{osc}}$.  
The CBT generator   
 (\ref{ConformalBridge1}) allows  
  to  define 
the pseudo-Hermitian inner
 product 
\begin{eqnarray}\label{pseudoscalar}
(\phi_{n_{1}},\phi_{n_{2}}) :=
\bra{\phi_{n_{1}}}\hat{\Theta}\ket{\phi_{n_{2}}}=\bra{\psi_{n_{1}}}\ket{\psi_{n_{2}}}=\delta_{n_1n_2}\,,\qquad 
\hat{\Theta}=  \hat{\Theta}^\dagger=
\hat{\mathfrak{S}}
^{2}\,,
\end{eqnarray}
which reduces to  a usual scalar product for orthonormalized eigenfunctions of
  the harmonic oscillator.
   
Using  definition (\ref{pseudoscalar}), we introduce the pseudo-Hermitian conjugation  $\hat{O}^{'\dagger}$ of 
an operator $\hat{O}$,
\begin{equation}
(\phi_{n_{1}},\hat{O}\phi_{n_{2}})=
(\hat{O}^{'\dagger}\phi_{n_{1}},\phi_{n_{2}})\,. 
\end{equation} 
By developing the left hand side of this equation, one gets 
\begin{eqnarray}
(\phi_{n_{1}},\hat{O}\phi_{n_{2}})&=&
\bra{\phi_{n_{1}}}\hat{\Theta}\hat{O}\ket{\phi_{n_{2}}}
 =\bra*{\hat{O}^\dagger \hat{\Theta}\phi_{n_{1}}}\ket*{\phi_{n_{2}}}=
  \bra*{\hat{\Theta}^{-1}\hat{O}^\dagger \hat{\Theta}\phi_{n_{1}}}\hat{\Theta} \ket*{\phi_{n_{2}}}\nonumber\\
  &=&( \hat{\Theta}^{-1}\hat{O}^\dagger \hat{\Theta}\phi_{n_{1}},\phi_{n_{2}})\,,
\end{eqnarray}
from where we identify 
\begin{eqnarray}
\label{pseudeHercon}
\hat{O}^{'\dagger}=\hat{\mathfrak{S}}^{-2}\hat{O}^{\dagger}\hat{\mathfrak{S}}^2\,. 
\end{eqnarray}
The application of this to
the 
Hermitian operators ($\hat{x}$, $\hat{H}$, $\hat{K}$) and anti-Hermitian operators
 $i\hat{p}$ and  $i\hat{D}$
 gives us the (\textit{a priori}) contra-intuitive relations 
 with respect to the pseudo-Hermitian conjugation,
\begin{eqnarray}
&
 (\frac{i}{\sqrt{m\omega\hbar}}\hat{p})^{'\dagger}=\sqrt{\frac{m \omega}{\hbar }}\hat{x}\,,
\qquad 
(\sqrt{\frac{m\omega}{\hbar}}\hat{x})^{'\dagger}=\frac{i}{\sqrt{m\omega\hbar}}\hat{p}\,,
&\\&
 (-\frac{1}{\omega\hbar} \hat{H})^{'\dagger}=\frac{\omega}{\hbar}\hat{K}\,,
\qquad 
(\frac{\omega}{\hbar}\hat{K})^{'\dagger}=
-\frac{1}{\omega\hbar} \hat{H}\,,\qquad 
(i\hat{D})^{'\dagger}=i\hat{D}\,,&
\end{eqnarray}
which
are just the analogues of the relations 
$\hat{a}^\pm=(\hat{a}^\mp)^\dagger$, 
$\hat{J}_\pm=(\hat{J}_\mp)^\dagger$
and $\hat{J}_0=(\hat{J}_0)^\dagger$
for the harmonic oscillator.  
Thus, 
the $\mathcal{PT}$-symmetric
  Hamiltonian (\ref{Swanson2iD}) is pseudo-Hermitian with
  respect to the scalar product (\ref{pseudoscalar}).
 Additionally,  from (\ref{CBT1}) 
  one can also see  that 
   our conformal bridge transformation generated by $\hat{\mathfrak{S}}$ effectively 
   is the eighth-order root  of the identity transformation, see 
  Ref. \cite{InzPly9} as well.

Having described the one-dimensional version of our system,
let us now turn  to its multidimensional generalizations. 
Before this, we note that, up to an imaginary factor $i$, the first-order operator 
$\hat{H}_\mathcal{PT}$ looks like the Berry-Keating Hamiltonian \cite{BerKea,BerKea+}, 
which has been considered in the literature in the context of the relation between the Riemann 
hypothesis on  zeros of zeta-function and quantum mechanics 
\cite{Connes,SierTow,BenderRieman,BenderRieman2,Sierra}. 
We will return to this issue in the last section.

As it was shown in \cite{ERIHO}, it is simple to generalize this picture to
$d$-dimensional systems. 
To do that, we just consider the changes (summation over
 a  repeated index is assumed)
\begin{eqnarray}
&\label{NotationI}
\hat{x}\rightarrow \hat{x}_{j}\,,\qquad\hat{p}\rightarrow \hat{p}_{j}=
-i\hbar\frac{\partial}{\partial x_{j}}\qquad {j}=1,\ldots, d\,,
&\\&
\label{HKDind}
\hat{H}=\frac{1}{2m}\hat{p}_{j}\hat{p}_{j}\,,\qquad
\hat{K}=\frac{m}{2}\hat{x}_{j}\hat{x}_{j}\,,\qquad
\hat{D}=\frac{1}{4}\{\hat{x}_{j},\hat{p}_{j}\}\,,&
\\
&
\hat{J}_0=\frac{1}{2\omega\hbar}\hat{H}_{\text{osc}}\,,\quad
\hat{J}_\pm=\frac{1}{2}\hat{a}_{j}^{\pm}\hat{a}_{j}^{\pm}\,,\quad
\hat{H}_{\text{osc}}=\hbar\omega(\hat{a}_{j}^+\hat{a}_{j}^-+\frac{d}{2})\,,\quad
\hat{a}_{j}^\pm=\sqrt{\frac{m\omega}{2\hbar}}(\hat{x}_{j}
\mp\frac{\hbar}{m\omega}\frac{\partial}{\partial x_{j}})\,,\qquad 
\label{NotationF}
&
\end{eqnarray}
and construct the CBT generator
(\ref{ConformalBridge1}) using $\hat{H}$ and 
$\hat{K}$ defined in
(\ref{HKDind}).
 The mapping is like (\ref{CBT1}), but with the
inclusion of the index ${j}$ when it corresponds. 
Then for the $d$-dimensional $\mathcal{PT}$-symmetric Hamiltonian 
$\hat{H}_\mathcal{PT}=2i\omega\hat{D}$, any manifestly scale invariant operator 
$(\hat{x}_{j})^{l}(\hat{p}_{k})^{l}$,
where $l$ is any non-negative integer,
is an integral of motion. At the same time,
 the angular momentum tensor 
\be
\hat{M}_{jk}=\hat{x}_{j}\hat{p}_{k}-
\hat{x}_{k}\hat{p}_{j}
\ee
 is invariant under the CBT
since conformal generators (\ref{HKDind}) are rotationally invariant.

 In the following sections, we generalize the described picture to generate
the one- and two-dimensional supersymmetric systems.

\section{CBT and one-dimensional supersymmetric systems}
\label{SecSUSYCBT1d}
The aim
of this section is to 
 construct supersymmetric 
generalizations of the 
$\mathcal{PT}$-symmetric Hamiltonian (\ref{Swanson2iD}) 
and its bosonic and fermionic integrals of motion.
By means of an appropriate modification of the CBT 
we associate the obtained models with supersymmetric extensions of 
the
one-dimensional harmonic oscillator.
The developed construction will be used then
to study the higher dimensional case. 

\subsection{Extension with $N$ subsystems}
\label{SubSecNxN}
A generalized one-dimensional $N$-supersymmetric system
can be given by the $N\cross N$ matrix Hamiltonian  operator
and the set of intertwining operators,
\begin{eqnarray}
\label{NSUSY}
\hat{\mathcal{H}}=\text{diag}\,(\hat{H}_N,\hat{H}_{N-1},\ldots,\hat{H}_1)\,,\qquad 
\hat{A}_q\hat{H}_q=\hat{H}_{q+1}\hat{A}_q\,,\qquad q=1,\ldots,N-1\,.
\end{eqnarray}
In the conventional case with the 
second order matrix Schr\"odinger operator $\hat{\mathcal{H}}$,
the chain of subsystems' Hamiltonians  
and the 
intertwining operators $\hat{A}_q$ 
can be obtained from a systematic procedure known as the Darboux
transformation \cite{MatSal,Cooper}. The simplest example corresponds to the case 
\begin{eqnarray}\label{Hi=H}
&
\hat{H}_1  = \ldots =\hat{H}_N = \hat{H}\,,\qquad
\hat{A}_{q} = \frac{1}{\sqrt{2m}}\hat{p}\,, 
&
\end{eqnarray}
where $\hat{H}$ is the one-dimensional 
free particle Hamiltonian operator 
introduced in (\ref{so21gen}). 
The integrals of motion of  this system 
 are given by  constant matrices 
 $ E_{jk}$ with elements
 \begin{eqnarray}
 &\label{Eij}
 (E_{{jk}})_{lm}=\delta_{jl}\delta_{km}\,, 
&
 \end{eqnarray}
and the matrix first order differential operators
  $\hat{p}E_{ij}$.
Identification of the integrals 
as even and odd generators of the corresponding superalgebra
depends on the  choice
of the grading operator, 
and we will return to this point later.  
At the same time, the diagonal operators, 
\begin{eqnarray}
\label{So(21)superGen}
\hat{\mathcal{H}}=\hat{H}\I_{N\cross N}\,,\qquad 
\hat{\mathcal{D}}=\hat{D}\I_{N\cross N}\,,\qquad 
\hat{\mathcal{K}}=\hat{K}\I_{N\cross N}\,,
\end{eqnarray}
can be identified as generators 
of the $\mathfrak{so}(2,1)$  algebra, where $\I_{N\cross N}$ is the unit matrix of order $N$.

Let us take now the first order differential matrix  operator
\begin{eqnarray}
&\label{PTHNXN}
\hat{\mathscr{H}}=2i\omega\hat{\mathcal{D}}+\hbar\omega\vartheta\,,\qquad
\vartheta_{jk}=(N-{j}-\frac{1}{2})\delta_{jk}\,,
\qquad {j},{k}=1,\ldots,N\,,
&
\end{eqnarray}
as the $\mathcal{PT}$-symmetric Hamiltonian. 
The constant diagonal matrix $\vartheta$  
is included here
 to obtain 
the $N$-supersymmetric extension of the harmonic oscillator system 
after 
applying the CBT operators
\be
\label{CBT3x3}
\hat{\mathscr{S}}=
\hat{\mathfrak{S}}\I_{N\cross N}\,,
\qquad
\hat{\mathscr{S}}^{-1}=\hat{\mathfrak{S}}^{-1}\I_{N \cross N}\,,\qquad
\hat{\Theta}=\hat{\mathscr{S}}^{2}  \,,
\ee
to the Hamiltonian (\ref{PTHNXN}) \footnote{Such a system is obtained
 by choosing $\hat{H}_1=\hat{H}_{\text{osc}}-\frac{1}{2}\hbar\omega=\hbar\omega\hat{a}^+\hat{a}^-$ and 
$\hat{A}_q=\hat{a}^{-}$ in (\ref{NSUSY}) for all $q$.
As a consequence, $\hat{H}_{N-q+1}=\hat{H}_{\text{osc}}+\hbar\omega(N-q-\frac{1}{2})$, and 
(\ref{NSUSY}) 
coincides with (\ref{Superoscillator}). },
\begin{eqnarray}
\label{Superoscillator}
\hat{\mathscr{S}}(\hat{\mathscr{H}})\hat{\mathscr{S}}^{-1}=
\hat{H}_{\text{osc}}\I_{N\cross N}+\hbar\omega\vartheta\,.
\end{eqnarray}

It is convenient to represent 
$\vartheta$  in the equivalent form 
\begin{eqnarray}
&
\vartheta=(\frac{N}{2}-1)\I_{N\cross N}+\tau\,,\qquad \tau=\sum_{f=1}^{N-1}c_{f}h_{f}\,,
&
\end{eqnarray}
where $h_{f}$ are the
 $N-1$ diagonal traceless matrices 
normalized by the inner product
Tr$(h_{{j}}\cdot h_{k})=2\delta_{jk}$, and 
$c_{f}$ are the real constants, 
$c_{f}=\frac{1}{2}\text{Tr}(h_{f}\vartheta)$. 
The eigenstates of the operator (\ref{PTHNXN}), which diagonalize
the 
$h_{f}$
matrices as well,  
are given by 
\begin{eqnarray}
&\label{statescone}
\Phi_{n,k}(x)=\phi_{n}(x)e_{N-\check{\imath}}^{T}\,,\qquad
e_{{j}}=(\underbrace{0,\ldots,0,1}_{{j}},\underbrace{0,\ldots,0}_{N-{j}
 })\,,
&\\&
\check{\imath}=0,\ldots,N-1\,,\quad n=0,1,\ldots,
&
\end{eqnarray}
where the functions $\phi_n(x)$ correspond to
(\ref{statesofsiD}), and $e_{N-\check{\imath}}^{T}$ 
indicates
  the transposition of the  unit
  vectors $e_{N-\check{\imath}}$. 
These states satisfy the eigenvalue equation 
\begin{eqnarray}
\hat{\mathscr{H}}\Phi_{n,\check{\imath}}(x)=E_{s}\Phi_{n,k}(x) \,, \qquad
E_{s}=\hbar\omega s\,,\qquad
s=n+\check{\imath}\,,
\end{eqnarray}
from where we note that energy levels 
$E_{s}$ with $s=0,1,\ldots N-1$ have degeneracy equal to $s+1$. 
Degeneracy of all other states with $s\geq N$ is  equal to $N$.

Having the $\mathcal{PT}$-symmetric Hamiltonian
(\ref{PTHNXN}) as well as its energy levels and eigenstates, 
we
identify now  the structure of its integrals of motion, and    
determine  their corresponding bosonic or fermionic nature. 
For this we note that the traceless matrices $h_f$ can be considered as 
generators of the Cartan sub-algebra  $\mathfrak{h}$  of 
$\mathfrak{su}(N)$.  The remaining $N(N-1)$ generators of  this Lie algebra
correspond to the constant matrices $E_{jl}$ with ${j}\not=l$  
introduced in (\ref{Eij}).
To find their commutation relation with our Hamiltonian, we just need to compute
\begin{eqnarray}
\label{HcomE}
[\tau, E_{jl}]=[\vartheta, E_{jl}]=(l-j)E_{jl}\,,
\end{eqnarray} 
from where we revel the nature of these \emph{dynamic} integrals
 as ladder operators in the second index of eigenfunctions, 
 $
E_{jl}\Phi_{n,\check{\imath}}=\Phi_{n,\check{\imath}+j-l}$.
Since $\tau\in \mathfrak{h}$,   
relation  (\ref{HcomE})
corresponds to the definition of a root vector of the corresponding simple Lie algebra 
of the Cartan series $A_{N-1}$. 
Then, for  $j>k$ we have the $N(N-1)$ 
\emph{true} integrals
\begin{eqnarray}
&\label{higherOrderInte}
\hat{I}_-^{(k,j)}=(\frac{i}{\sqrt{\hbar\omega m}}\hat{p})^{j-k}E_{kj}\,,\qquad 
\hat{I}_+^{(k,j)}=(\sqrt{\frac{m\omega}{\hbar}}\hat{x})^{j-k}E_{jk}\,,\quad
 j>k\,,  \qquad
[\hat{\mathscr{H}},\hat{I}_\pm^{(k,j)}]=0\,,\qquad
&\\&
\hat{I}_-^{(k,j)}\Phi_{n,\check{\imath}}\sim\Phi_{n-j+k,\check{\imath}+k-j}\,,\qquad
\hat{I}_+^{(k,j)}\Phi_{n,\check{\imath}}\sim\Phi_{n+j-k,\check{\imath}-k+j}\,,\qquad
&
\end{eqnarray}  
which satisfy the pseudo-Hermitian conjugation relations 
$(\hat{I}_\pm^{(k,j)})^{'\dagger}=\hat{I}_\mp^{(k,j)}$
 in the sense of (\ref{pseudeHercon}). 
Generally, 
these integrals generate 
 a nonlinear polynomial superalgebra.

In order to determine the fermionic or bosonic nature of dynamic and true integrals $E_{jl}$ and
(\ref{higherOrderInte}), we have to select
the grading operator $\Gamma$
such that $\Gamma^2=\I_{N\cross N}$, $[\hat{\mathscr{H}},\Gamma]=0$.
Here, we restrict ourselves to the case of even $N$, 
and for the sake of definiteness choose 
 $\Gamma$  
in the form of the diagonal traceless matrix
\begin{eqnarray}
\label{HigherOrderGrading}
\Gamma=\text{diag\, }(1,-1,\ldots, 1,-1)\,,
\end{eqnarray} 
 being 
 a linear combination of generators of the $\mathfrak{su}(N)$
 Cartan sub-algebra $\mathfrak{h}$. 
With this choice of the $\Z_2$-grading operator, we find 
  \begin{eqnarray}
&\label{BosFerm}
[\Gamma,E_{jk}]=0,\quad
 {j}-k=0\, (\text{mod}\,\,2)\,,\qquad
\{\Gamma,E_{jk}\}=0,\quad
 {j}-k=1\, (\text{mod}\,\,2)\,.\quad
 &
\end{eqnarray} 
 So,
the  first  subset of dynamic integrals  $E_{jk}$ in (\ref{BosFerm})
is identified as  even (bosonic)
generators of the superalgebra, while their second half 
 is identified as odd (fermionic) operators of the system (\ref{PTHNXN}).
 Coherently with this, the even and odd nature of 
 the true integrals   (\ref{higherOrderInte})
 is  identified.

We notice here that the permutation of the diagonal elements in 
(\ref{HigherOrderGrading}) provides us with alternative 
choices for the grading operator $\Gamma$. This results in the 
change of identification  of the even and  odd nature of the generators $E_{jl}$
and nontrivial integrals (\ref{higherOrderInte}).
As a consequence, the concrete form of the corresponding 
(nonlinear) superalgebra of the system will be changed.

Finally, last but not least,
from the free particle generators of translation 
and Galilean boost we identify the first order bosonic ladder operators of the system, 
\begin{eqnarray}
&\label{Amathscr}
\hat{\mathscr{A}}^-=\frac{i}{\sqrt{m\omega\hbar}}\hat{p}\,\I_{N\cross N}
\qquad 
\hat{\mathscr{A}}^+=\sqrt{\frac{m\omega}{\hbar}}\hat{x}\,\I_{N\cross N}\,,&\\&
[\hat{\mathscr{H}},\hat{\mathscr{A}}^\pm]=\pm \hbar\omega \hat{\mathscr{A}}^\pm \,,\qquad 
[\hat{\mathscr{A}}^-,\hat{\mathscr{A}}^+]=1\,.
&
\end{eqnarray}
Their square yields us 
 the second order ladder operators 
of the centrally extended
 $\mathfrak{sl}(2,\R)$ 
 conformal symmetry of the model  
 (\ref{PTHNXN}),
\begin{eqnarray}
\label{Jmathscr}
&\hat{\mathscr{J}}_0=\frac{1}{2\hbar\omega}\hat{\mathscr{H}}\,,\qquad
\hat{\mathscr{J}}_-=-\frac{1}{\hbar\omega}\hat{\mathcal{H}}\,,\qquad 
\hat{\mathscr{J}}_+=\frac{\omega}{\hbar}\hat{\mathcal{K}}\,,
&\\
&\label{ExtendedSL2R}
[\hat{\mathscr{J}}_0,\hat{\mathscr{J}}_\pm]=\pm \hat{\mathscr{J}}_\pm\,,\qquad
[\hat{\mathscr{J}}_-,\hat{\mathscr{J}}_+]=2\hat{\mathscr{J}}_0-\vartheta\,,\qquad
[\vartheta,\hat{\mathscr{J}}_0]=[\vartheta,\hat{\mathscr{J}}_\pm]=0\,.
&
\end{eqnarray}
Under the CBT, these generators map into the second and first order ladder operators of the superexteded system
(\ref{Superoscillator}) in accordance
with (\ref{CBT1}).

\vskip0.1cm

Let us  
consider now two concrete
examples. 
\begin{itemize}
\item
\textit{The
$N=2$ case}.
\end{itemize}
In this 
simplest case,
$\vartheta=\tau=\frac{1}{2}\sigma_3$, $\Gamma=\sigma_3$ and 
$
E_{12}=\frac{1}{2}\sigma_+\,,$
$E_{21}=\frac{1}{2}\sigma_-\,,$
$\sigma_\pm=\sigma_1\pm i\sigma_2$.  
The fermionic operators 
\begin{eqnarray}
&
\hat{Q}_-=\sqrt{\hbar\omega}\hat{I}_{12}=\frac{1}{2}\frac{i}{\sqrt{m}}\hat{p} \sigma_+\,,\qquad
\hat{Q}_+=\sqrt{\hbar\omega}\hat{I}_{21}=\frac{1}{2}\sqrt{m\omega^{2}}\hat{x}\sigma_-
&
\end{eqnarray}
are the  true integrals of motion, 
which  satisfy the $\mathcal{N}=2$ Poincar\'e superalgebra 
\begin{eqnarray}
&
[\hat{\mathscr{H}},\hat{Q}_\pm]=0\,,\qquad
\{\hat{Q}_-,\hat{Q}_+\}=\hat{\mathscr{H}}\,.
&
\end{eqnarray}
The ground state of the system is 
non-degenerate,
 and all the excited states  
 are doubly degenerate.
 
The inclusion of  the integral $\sigma_3$
 and second order ladder operators (\ref{Jmathscr}) extends 
  the Poincar\'e supersymmetry  to the $\mathfrak{osp}(2,2)$ superalgebra, 
  generating in the process the dynamical integrals 
 $\hat{S}_-=\frac{1}{2}\frac{i}{\sqrt{m}}\hat{p} \sigma_-$ and
 $\hat{S}_+=\frac{1}{2}\sqrt{m\omega^{2}}\hat{x}\sigma_+\,.$
By adding  the first order ladder operators (\ref{Amathscr}), one gets  
 the super-Schr\"odinger symmetry, where the  fermionic ladder 
 operators $\sigma_\pm$  are also included.
Accordingly, the 
 application of the CBT to this $\mathcal{PT}$-symmetric model yields 
the usual harmonic  super-oscillator system
 together with its entire symmetry superalgebra \cite{InzPly1}.
 
This  model can be compared with supersymmetric 
generalizations of the Swanson model already treated in the literature by
changing $\omega^2\rightarrow \Omega_{\alpha,\beta}^2=\omega^2-4\alpha\beta>0$
and considering the system of units $m=\omega=\hbar=1$ for a moment. 
By  applying  the similarity transformation 
$\hat{\mathscr{T}}_a(\hat{\mathscr{H}})\hat{\mathscr{T}}^{-1}_a$ to (\ref{PTHNXN}) with $N=2$,
where
$\hat{\mathscr{T}}_a= \hat{T}_a^{-1}\hat{\mathfrak{S}}_{\Omega_{\alpha,\beta}}\I_{2\cross 2}$,
and 
$\hat{T}_a$ are defined in Appendix \ref{AppLabel} in dependence on whether 
$\alpha+\beta < 1$ ($a=1$)
or $\alpha+\beta>-1$ ($a=2$),  
one gets 
\begin{eqnarray}
&
\hat{H}_{SS}=\text{diag}\,(\hat{H}_+,\hat{H}_-)\,,\qquad
\hat{H}_\pm=\hat{a}^\mp \hat{a}^\pm +\alpha(\hat{a}^-)^2+\beta (\hat{a}^+)^2\,.
&
\end{eqnarray}
The supersymmetric partners $\hat{H}_\pm$ have the 
form of those in the generalized supersymmetric Swanson models discussed in 
Ref. \cite{SUSYSwanson1,SUSYSwanson2} with superpotential $\hat{W}=\hat{x}$. 
The intertwining operators are obtained 
from the action of the corresponding similarity 
transformation on $\hat{x}$ and $\hat{p}$.  
\begin{itemize}
\item
\textit{The $N=4$ case}.
\end{itemize}

For this case,
the constant matrix $\vartheta$ 
takes the form 
\begin{eqnarray}
&
\vartheta=\frac{1}{2}\text{diag}\,(5,3,1,-1)=\frac{1}{2}\I_{4\cross 4}+\frac{1}{2}h_1+\frac{\sqrt{3}}{2}h_2+\frac{\sqrt{6}}{2}h_3\,,
& 
\end{eqnarray}
where the three  $\mathfrak{su}(4)$  Cartan  integrals 
are given by 
$h_1=E_{11}-E_{22}$,
$h_2=\frac{1}{\sqrt{3}}(E_{11}+E_{22}-2E_{33})$
and 
$h_3=\frac{1}{\sqrt{3}}(E_{11}+E_{22}+E_{33}-3E_{44})$. 
The dynamical integrals 
 $E_{jk}$ with ${j}, k=1,2,3,4$, ${j}\neq k$,   
 are used to construct the  true integrals 
$I_{-}^{({j},k)}$ and $I_{+}^{({j},k)}$.
The $\Z_2$-grading operator  
\begin{eqnarray}
&\label{GammaN=4}
\Gamma=\text{diag}\,(1,-1,1,-1)=h_1-\frac{1}{\sqrt{3}}h_2+
\sqrt{\frac{2}{3}}h_3 &
\end{eqnarray}
identifies the set of eight true  integrals of motion 
$\mathcal{F}=(\hat{I}_\pm^{(1,2)},\hat{I}_\pm^{(1,4)},\hat{I}_\pm^{(2,3)},\hat{I}_\pm^{(3,4)}  )$
as   
fermionic generators (supercharges),
from which only  $\hat{I}_\pm^{(1,4)}$
are of order three, and the rest   
are of order one 
operators in $\hat{p}$ and $\hat{x}$.
Eight generators
  $\mathcal{B}=(\hat{\mathscr{H}}_\mathcal{PT},h_1,h_2,h_3,\hat{I}_\pm^{(1,3)},\hat{I}_\pm^{(2,4)})$ are the
 true  bosonic integrals of motion.  
The entire  superalgebra of this system 
 is nonlinear due to the presence of the higher order operators. 
The ground state is singlet, 
while the first and the  second excited energy levels have degeneracies 
two and three, respectively.  Degeneracy of all  
other excited states is equal to  four.  

The inclusion of the first and second order ladder operators significantly increases the number of generators in this case, 
 and 
 essentially complicates the structure of the corresponding nonlinear superalgebra. 
 Instead of the detailed analysis of its structure, we just note that 
 the computation of  
the commutator between $\hat{I}_\pm^{(1,4)}$ and $\hat{\mathscr{A}}^\pm$ 
generates  the four odd dynamical integrals $\hat{\mathscr{A}}^{-} E_{14}$,
$\hat{\mathscr{A}}^{+} E_{41}$,
$(\hat{\mathscr{J}}^{-}) E_{14}$ and 
$(\hat{\mathscr{J}}^{+}) E_{41}$. 
Then, the inclusion 
of   the second order operators as well  extends and complicates further the 
nonlinear superalgebraic structure.

\subsection{Higher order $N=2$ superextension}
Let us consider again the case $N=2$ and introduce the generalized model 
\be
\label{PT-her1}
\hat{\mathscr{H}}_\gamma= 2i\omega \hat{\mathcal{D}}+\frac{1}{2}\gamma\hbar \omega \sigma_3+\frac{\omega\hbar}{2}(\gamma-1)
=\left(
\begin{array}{cc}
\hat{H}_{\mathcal{PT}}+\hbar\omega(\gamma-\frac{1}{2}) &0\\
0& \hat{H}_{\mathcal{PT}}-\frac{\hbar\omega}{2}
\end{array}
\right)\,,
\ee
where $\gamma$ is, in principle,  an arbitrary  numerical parameter.
Note that  we have a copy of $\hat{H}_{\mathcal{PT}}$ in each subsystem  
in (\ref{PT-her1}), but they occur with a 
relative  energy offset equal to $ \hbar\omega \gamma $. 
Taking into account this 
shift and the nature of the spectrum of $ \hat{H}_{\mathcal{PT}} $, 
governed by a non-negative
integer as it is shown in (\ref{statesofsiD}), the system $\hat{\mathscr{H}}_\gamma$
will be supersymmetric only when $ \gamma=l \in \Z $, and without loss of generality 
we assume in the following $l\geq 0$.

The application of the CBT 
produces
$\hat{\mathscr{S}}(\hat{\mathscr{H}}_l) \hat{\mathscr{S}}^{-1}=
\hat{\mathcal{H}}_l$, 
where 
\begin{eqnarray}
\label{SUSYosc}
&
\hat{\mathcal{H}}_l=\hat{H}_{\text{osc}}\I_{N\cross N}+\frac{1}{2}\hbar\omega l \sigma_3+\frac{\hbar\omega}{2}(l-1)=
\left(\begin{array}{cc}
\hat{H}_{\text{osc}} +\hbar\omega(l-\frac{1}{2}) & 0\\
0 &\hat{H}_{\text{osc}} -\frac{\hbar\omega}{2}
\end{array}
\right)\,.&
\end{eqnarray}
When $l=1$, operator $\hat{\mathcal{H}}_1$ takes the form of the usual super-extended 
harmonic oscillator system, 
while
for $l=2,\ldots$, 
(\ref{SUSYosc}) corresponds to a higher-order supersymmetric extension of the model 
which is
  characterized by nonlinear supersymmetry \cite{InzPly3}. Finally, 
 in the case $l=0$, one obtains the ``order zero" supersymmetric extension 
 \cite{KliPlyu,InzPly1}.

 The eigenstates of (\ref{PT-her1}) with $\gamma=l$ 
 are provided by equations (\ref{statescone}) with $N=2$.
However,
 it is convenient to present them in the basis 
\begin{eqnarray}
\label{stateground1}
&\Phi_{j}^{(0,l)}(x)=\left(\begin{array}{cc}
0\\
\phi_{j}(x)
\end{array}
\right)\,,\quad j=0,1\ldots,l-1\,, 
\quad
\Phi_{n+l}^{(\pm,l)}=
\frac{1}{\sqrt{2}}\left(\begin{array}{cc}
\phi_{n+l}(x)\\
\pm \phi_{n}(x)
\end{array}
\right)\,,
&
\label{stateground2}
\end{eqnarray}
which
 satisfy the eigenvalue equation 
  \begin{eqnarray}
&
\hat{\mathscr{H}}_l\Phi_{n}^{(\lambda,l)}(x)=\hbar\omega n\Phi_{n}^{(\lambda,l)}(x) \,,\quad
\lambda=\pm,0\,.
&
\end{eqnarray}
From here it is clear that the value of $l$  is equal to  the number of 
non-degenerate, 
 singlet states in  
the system.
Note that in the case $l=0$, we  just 
have two copies of the same operator $\hat{H}_\mathcal{PT}$, 
and  all energy levels $\hbar\omega n$ in its 
 spectrum 
are doubly  degenerate, with corresponding eigenstates 
given by
$\Phi_n^{(\pm,0)}$.  

Similar to the case $N=2$ analyzed in the previous section, 
the only true fermionic  integrals of motion of the
system (\ref{PT-her1}) with $l\geq 2$ are the 
two supercharges,
but in this case their structure is given in terms of the
higher order operators,
\begin{eqnarray}
&\label{superchargesHigherorderos}
\hat{Q}_{l}^{-}=\frac{1}{2}(\frac{i}{\sqrt{\hbar\omega m}}\hat{p})^{l}\sigma_+\,,\qquad 
\hat{Q}_{l}^{+}=\frac{1}{2}(\sqrt{\frac{m\omega}{\hbar}}\hat{x})^{l}\sigma_-\,, \qquad
[\hat{\mathscr{H}}_l,\hat{Q}_{l}^{\pm}]=0\,.
&
\end{eqnarray} 
These integrals 
are mapped  by the CBT into the higher order supercharges of the 
$N=2$ super-extended harmonic oscillator system,
\begin{eqnarray}
&\label{CBTonQ}
\hat{\mathscr{S}}(\hat{Q}_{l}^{\pm})\hat{\mathscr{S}}^{-1}=\hat{\mathcal{Q}}_{l}^{\pm}\,,\qquad 
\hat{\mathcal{Q}}_{l}^{\pm}= \frac{1}{2}(\hat{a}^\pm)^l \sigma_\mp\,. &
\end{eqnarray}

It is convenient to present operators (\ref{superchargesHigherorderos}) in the pseudo-Hermitian basis 
(in the sense of (\ref{pseudeHercon}))
 \begin{eqnarray}
&\label{SuperChargesPT1}
\hat{Q}_{l}^{(1)}=
\hat{Q}_{l}^{-} + \hat{Q}_{l}^{+}
=
\left(\begin{array}{cc}
0 & (\frac{i}{\sqrt{\hbar\omega m}}\hat{p})^{l} 
\\
  (\sqrt{\frac{m\omega}{\hbar}}\hat{x})^{l} & 0
\end{array}\right)\,,\qquad
\hat{Q}_{l}^{(2)}=-i\sigma_3 \hat{Q}_{l}^{(1)}\,,
&
\end{eqnarray} 
since in this way we have their action on  
eigenstates (\ref{stateground1}) 
in  a
simple form 
\begin{eqnarray}
&
\hat{Q}_{l}^{(1)}\Phi_{n+l}^{(\pm,l)}=\pm \sqrt{(n+1)
\ldots(n+l)}\Phi_{n+l}^{(\pm,l)}\,,\qquad
\hat{Q}_{l}^{(a)}\Phi_{j}^{(0,l)}=0\,, \quad 
&\\ &
\hat{Q}_{l}^{(2)}\Phi_{n+l}^{(\pm,l)}=\mp i\sqrt{(n+1)
\ldots(n+l)}\Phi_{n+l}^{(\mp,l)}\,.&
\end{eqnarray}
As the odd operators are  the
true integrals, and  together  with  the 
corresponding Hamiltonian annihilate the singlet  ground state  $\Phi_{0}^{(0,l)}$
(along with all other singlet states),
we conclude that for a given $l\geq 2$,
 the system (\ref{PT-her1}) is in the exact supersymmetric phase.  
 The anti-commutator  of
 supercharges 
 produces a polynomial of order $l$ in the corresponding Hamiltonian operator,  
\begin{eqnarray}
&
\{
\hat{Q}_{l}^{(a)},\hat{Q}_{l}^{(b)}\}=2\delta_{ab}\Pi_{j=0}^{l-1}\left(\frac{1}{\hbar\omega}\hat{\mathscr{H}}_l
-j\right)\,.
&
\end{eqnarray}

Let us extend the picture by looking for the  dynamical,
with respect to the Hamiltonian $\hat{\mathscr{H}}_l$,  integrals of motion.
These models also have the first and the second order ladder operators 
(\ref{Amathscr}) and (\ref{Jmathscr}). 
The  only difference 
appears here in the second commutation relation in (\ref{ExtendedSL2R}), where $\vartheta$
has to
be replaced by the matrix operator $\frac{1}{2}(l (\sigma_3+1)-1)$. 
Together with them, we have the R-symmetry generator $\hat{\mathcal{R}}=\frac{\hbar}{2}\sigma_3$
and the higher order dynamical fermionic operators
\begin{eqnarray}
&\label{S+-l}
\hat{S}_l^{-}= \frac{1}{2}(\frac{i}{\sqrt{\hbar\omega m}}\hat{p})^{l}\sigma_-\,,\qquad 
\hat{S}_l^{+}= \frac{1}{2}(\sqrt{\frac{m\omega}{\hbar}}\hat{x})^{l}\sigma_+\,,\qquad
\hat{\Sigma}_\pm=\frac{1}{2}\sigma_\pm\,.
&
\end{eqnarray}
Some of the superalgebraic  relations that these operators satisfy 
are 
 \begin{eqnarray}
 &
 \{
\hat{S}_{l}^{\pm},\hat{S}_{l}^{\mp}\}=\Pi_{j=0}^{l-1}\left(\frac{1}{\hbar\omega}\hat{\mathscr{H}}_l-\frac{2 l}{\hbar}\hat{\mathcal{R}}
-j\right)\,,\qquad 
\{\hat{S}_{l}^{\pm},\hat{S}_{l}^{\pm}\}=0&\\
&[\hat{\mathcal{R}},\hat{Q}_l^{\pm}]=\pm \hbar  \hat{Q}_l^{\pm} \,,\qquad 
[\hat{\mathcal{R}},\hat{S}_l^{\pm}]=\mp \hbar\hat{S}_l^{\pm}\,,&\\&
[\hat{\mathscr{H}}_l,\hat{S}_l^{\pm}]=\mp 2 \hbar\omega l \hat{S}_l^{\pm}\,,\qquad 
[\hat{\mathscr{H}}_l,\sigma_\pm]= \pm l\hbar \omega\sigma_\pm\,,& \\
&
\{\hat{\Sigma}_+,\hat{S}_l^{+}\}=\{\hat{\Sigma}_-,\hat{Q}_l^{+}\}=(\hat{\mathscr{A}}^+)^{l}\,,\qquad 
\{\hat{\Sigma}_-,\hat{S}_l^{-}\}=\{\hat{\Sigma}_+,\hat{Q}_l^{-}\}=(\hat{\mathscr{A}}^{-})^{l}
\,,
&\\&
\{\hat{S}_{l}^{+},\hat{Q}_{l}^{+}\}=(2\hat{\mathscr{J}_+})^{l}\,,\qquad 
\{\hat{S}_{l}^{-},\hat{Q}_{l}^{-}\}=(2\hat{\mathscr{J}_-})^{l}\,.&
\end{eqnarray}

The application of the CBT to
the bosonic operators $\hat{\mathscr{J}}_\pm$ and $\hat{\mathscr{A}^\pm}$ works as we explained in 
Sec. \ref{SubSecNxN}. 
On the other hand, the matrix operators 
$\hat{\mathcal{R}}$ and $\hat{\Sigma}_\pm$
 are invariant under this mapping. 
The transformation of the supercharges $\hat{Q}_l^\pm$ is  shown  
 in Eq.  
 (\ref{CBTonQ}), and for  the dynamical fermionic integrals (\ref{S+-l})
we have 
\begin{eqnarray}
&
\hat{\mathscr{S}}(\hat{S}_{l}^\pm)\hat{\mathscr{S}}^{-1}=\hat{\mathcal{S}}_{l}^\pm\,,\qquad
\hat{\mathcal{S}}_{l}^\pm=\frac{1}{2}(\hat{a}^\pm)^{l}\sigma_\pm\,.
&
\end{eqnarray}

As noted 
in the previous subsection,
the transformed generators produce the super-Schr\"odinger symmetry in the case $l=1$, 
that completely describes the usual super-oscillator system, 
including the spectrum generating ladder operators \cite{InzPly1}.
 In the case $l\geq 2$, the described  set of generators   produces 
  a more complicated,  
 nonlinearly deformed extension of the superconformal algebra, which  involves 
  some  additional higher order generators, 
see Ref. \cite{InzPly3}. 
 
The ideas applied in this subsection  
 can 
 be extended to the case of arbitrary number of
 subsystems by  taking 
 $\vartheta$ instead of $\sigma_3$ in (\ref{PT-her1}).
 The appearance of even more complicated nonlinear superalgebraic
 structure is expected in this case.

%
%%
%%%
%%%%
%%%%%
%%%%%%
%%%%%%%
%%%%%%%%
%%%%%%%%%
%%%%%%%%%%
%%%%%%%%%%%
%%%%%%%%%%%%
%%%%%%%%%%%%%
%%%%%%%%%%%%%%
%%%%%%%%%%%%%%%

\section{CBT and two-dimensional supersymmetric systems}
\label{SecSUSYCBT2d}
This section is devoted to the
study of 
supersymmetric extensions of two-dimensional
generalizations of the Swanson model
(\ref{Swanson2iD}). 
In subsection \ref{SubsecNonRecLim},  
an $\mathcal{N}=4$ superconformal invariant $2D$ free particle
model  is obtained from the non-relativistic limit of the
Dirac and Klein-Gordon equations in a  $(2+1)$-dimensional 
Minkowski spacetime. 
In subsection \ref{SubsecLandau}, we use the dynamic
 symmetry  generators of
the resulting model to 
 construct a two-dimensional non-relativistic $\mathcal{PT}$-invariant   
 supersymmetric system. The latter
 model
 is related to the supersymmetric 
 Landau problem via the rotationally invariant CBT. 
Two different extensions of  the supersymmetric 
 Landau problem are obtained then
 in subsections 
\ref{SubSecAharonov-Bohm} and \ref{SubSecERIHO}. 
The first of them corresponds to
the inclusion of the Aharonov-Bohm flux. 
The second one is a modification that includes an angular momentum coupling term, and 
can be related with a supersymmetric extension of the exotic rotational invariant harmonic oscillator (ERIHO)
studied in \cite{ERIHO}. 

We employ  here the notation (\ref{NotationI})--(\ref{NotationF}) with $d=2$ for the two-dimensional 
 free  particle and 
 harmonic oscillator generators. 
  However,  instead of taking
  the Hermitian operators
  $(\hat{x}_i,\hat{p}_i)$ and $(\hat{a}_i^+,\hat{a}_i^-)$ as a basis, 
  it is convenient to use 
  \begin{eqnarray}
  &
  \hat{p}_\pm=\hat{p}_1\pm i\hat{p}_2\,,\qquad
  \hat{x}_\pm=\hat{x}_1\pm i\hat{x}_2\quad \text{and}
\qquad
\hat{b}_j^\pm=\frac{1}{\sqrt{2}}(\hat{a}_1^\pm \mp i(-1)^j 
 \hat{a}_2^\pm)\,.\label{b+b-}
& 
  \end{eqnarray}
These operators are related via  the CBT as follows, 
  \begin{eqnarray}
  &
  \hat{\mathfrak{S}}
  (
  \sqrt{\frac{\omega m}{2\hbar}}\hat{x}_+,
    \frac{i}{\sqrt{2m\omega\hbar}}\hat{p}_-,
  \sqrt{\frac{\omega m}{2\hbar}}\hat{x}_-,
  \frac{i}{\sqrt{2m\omega\hbar}}\hat{p}_+
  )
  \hat{\mathfrak{S}}^{-1}  =
  (\hat{b}_1^+,\hat{b}_1^-,\hat{b}_2^+,\hat{b}_1^-)\,,
  &
  \end{eqnarray}
implying the  pseudo-Hermitian conjugation relations 
  $(\sqrt{\frac{\omega m}{2\hbar}}\hat{x}_\pm)^{'\dagger}= \frac{i}{\sqrt{2m\omega\hbar}}\hat{p}_\mp\,.$
 The angular momentum operator  
 $\hat{M}_{12}=\hat{x}_1\hat{p}_2-\hat{x_2}\hat{p}_1=\hbar (\hat{b}_1^{+}\hat{b}_1^{-}-\hat{b}_2^{+}\hat{b}_2^{-})$ 
  is denoted as $\hat{p}_\varphi$.

\subsection{Taking a non-relativistic limit}
\label{SubsecNonRecLim}

 In a $(2+1)$-dimensional Minkowski spacetime, 
relativistic 
massive Dirac equation
\begin{eqnarray}
&\label{Dirac0}
i\hbar\frac{\partial}{\partial t}\Psi(t,\vx)=\hat{H}_\text{Dirac}\Psi(t,\vx)\,,\qquad 
\hat{H}_\text{Dirac}=
c(\beta mc+\alpha_i\hat{p}_i)\,,
\end{eqnarray}
takes the form
\begin{eqnarray}
&
\label{Dirac}
i\frac{\hbar}{c}\frac{\partial}{\partial t}\Psi(t,\vx)= (
\hat{\pi}
+mc\sigma_3)
\Psi(t,\vx)\,,\qquad 
\hat{\pi}=
\left(\begin{array}{cc}
0 & \hat{p}_-\\
\hat{p}_+& 0
\end{array}\right)\,,
&
\end{eqnarray}
under the
choice of representation 
$\beta=\sigma_3$, $\alpha_1=\sigma_1$
and $\alpha_{2}=\sigma_2$. 
Acting on equation (\ref{Dirac})
by $i\frac{\hbar}{c}\frac{\partial}{\partial t}$, 
one gets the Klein-Gordon equation
\begin{eqnarray}
\label{klein}&
\frac{-\hbar^2}{c^2}\frac{\partial^2}{\partial t^2}\Psi(t,\vx)=(m^2c^2+2m\hat{\mathcal{H}})\Psi(t,\vx)\,,\qquad 
\hat{\mathcal{H}}=\hat{H} \I_{2\cross 2}\,.&
\end{eqnarray} 
The non-relativistic limit of this spin-$1/2$ system  is obtained by 
taking 
the ansatz  
\be
\Psi(t,\vx)=e^{-\frac{i}{\hbar}(E+mc^2)t}\chi(\vx)\,,\qquad 
\chi(\vx)=\left(\begin{array}{c}
\chi_1(\vx)\\
\chi_2(\vx)
\end{array}\right)\,,
\ee
{and assuming   $E\ll mc^2$.
Then 
$(E+mc^2)^2\approx   
 m^2c^4+2mc^2E$, and  equation (\ref{klein}) takes the form of the stationary 
Schr\"odinger equation of a two-dimensional supersymmetric free particle model
\be\label{2DHfree}
\hat{\mathcal{H}}\chi(\vx)=E\chi(\vx)\,.
\ee
Here,  $\hat{\mathcal{H}}$ is  
a $2\cross 2$ matrix Hamiltonian, 
 whose diagonal elements are two copies of the usual two-dimensional non-relativistic 
 free particle Hamiltonian. 
 Taking 
 $\Gamma=\sigma_3$
 as the $\Z_2$-grading operator, the following operators
 \begin{eqnarray}\label{Int1}
&
\hat{\mathcal{D}}=\hat{D}\I_{2\cross 2}\,,\qquad 
\hat{\mathcal{K}}=\hat{K}\I_{2\cross 2}\,,\qquad 
\hat{\mathcal{L}}_0=\frac{1}{2}(\hat{p}_\varphi\I_{2\cross 2}+\frac{\hbar}{2}\sigma_3)\,,
&\\&
\hat{\mathcal{T}}_\pm=\frac{1}{2m}(\hat{\mathcal{P}}_\pm)^2\,,\qquad 
\hat{\mathcal{F}}_\pm=\frac{1}{2m}(\hat{\mathcal{X}}_\pm)^2\,,\qquad 
\hat{\mathcal{L}}_\pm=\frac{1}{2m}\hat{\mathcal{X}}_\pm \hat{\mathcal{P}}_\pm\,,
&\label{Int2}\\
&
\hat{\mathcal{P}}_\pm=\hat{p}_\pm\I_{2\cross 2}\,,\qquad 
\hat{\mathcal{X}}_\pm=m \hat{x}_\pm \I_{2\cross 2}\,,
\qquad 
\hat{\mathcal{M}}=m\I_{2\cross 2}\,,\label{Int3}
&
\end{eqnarray}
are identified as the even (bosonic) 
integrals of the matrix system 
  (\ref{2DHfree}).
Among them, only $\hat{\mathcal{L}}_0$, $\hat{\mathcal{P}}_\pm$ and $\hat{\mathcal{T}}_\pm$ are the true integrals
of motion that commute with the Hamiltonian operator, while $\hat{\mathcal{M}}$ is a central charge. 
All other generators are dynamical integrals of motion.  
Integrals  (\ref{Int1}), (\ref{Int2})
generate 
 the $\mathfrak{sp}(4)$ symmetry with the nonzero commutators
\begin{eqnarray}
&\label{Simetryal1}
[\hat{\mathcal{D}},\hat{\mathcal{H}}]=i\hbar\hat{\mathcal{H}}\,,
\qquad
[\hat{\mathcal{D}},\hat{\mathcal{K}}]=-i\hbar\hat{\mathcal{K}}\,,
\qquad 
[\hat{\mathcal{K}},\hat{\mathcal{H}}]=2i\hbar \hat{\mathcal{D}}\,,&\\&
[\hat{\mathcal{L}}_0, \hat{\mathcal{L}}_\pm]=\pm \hbar \hat{\mathcal{L}}_\pm\,,\qquad
[\hat{\mathcal{L}}_-, \hat{\mathcal{L}}_+]= 2 \hbar  \hat{\mathcal{L}}_0-\hbar\hat{\mathcal{R}}\,,\label{Simetryal2}
&\\&
\label{Euclidal}
[\hat{\mathcal{L}}_0, \hat{\mathcal{T}}_\pm]=\pm \hbar \hat{\mathcal{T}}_\pm\,,\qquad
[\hat{\mathcal{L}}_0,\hat{\mathcal{F}}_\pm]=\pm \hbar \hat{\mathcal{F}}_\pm\,,
&\\&
[\hat{\mathcal{H}},\hat{\mathcal{F}}_\pm]=-2i\hbar \hat{\mathcal{L}}_\pm\,,\qquad
[\hat{\mathcal{H}},\hat{\mathcal{L}}_\pm]=-i\hbar \hat{\mathcal{T}}_\pm\,,&\\&
[\hat{\mathcal{K}},\hat{\mathcal{T}}_\pm]=2i\hbar \hat{\mathcal{L}}_\pm\,,\qquad
[\hat{\mathcal{K}},\hat{\mathcal{L}}_\pm]=i\hbar \hat{\mathcal{F}}_\pm\,,&\\&
[\hat{\mathcal{D}},\hat{\mathcal{T}}_\pm]=i\hbar \hat{\mathcal{T}}_\pm \,,\qquad 
[\hat{\mathcal{D}},\hat{\mathcal{F}}_\pm]=-i\hbar \hat{\mathcal{F}}_\pm\,,&\\&
[\hat{\mathcal{F}}_\pm, \hat{\mathcal{T}}_\mp]=\pm 4\hbar(\hat{\mathcal{L}}_0\pm i\hat{\mathcal{D}})
\mp 2\hbar\hat{\mathcal{R}}\,,
&\\&
[\hat{\mathcal{L}}_\pm, \hat{\mathcal{T}}_\mp]=2i\hbar \hat{\mathcal{H}}\,,\qquad
[\hat{\mathcal{L}}_\pm, \hat{\mathcal{F}}_\mp]=2i\hbar \hat{\mathcal{K}}\,.
\label{Final?}&
\end{eqnarray} 
 The integral  $\hat{\mathcal{R}}=\frac{\hbar}{2}\sigma_3$ 
 commutes with all 
 generators (\ref{Int1})--(\ref{Int3}),  
 and as will be seen below, 
plays the role of the $R$-symmetry generator. 
Integrals  (\ref{Int3})
  produce an ideal Heisenberg subalgebra,
\begin{eqnarray}
&
[\hat{\mathcal{X}}_\pm,\hat{\mathcal{P}}_\mp]=2i\hbar\hat{\mathcal{M}}\,, &\label{Hei2}\\&
[\hat{\mathcal{H}},\hat{\mathcal{X}}_\pm]=-i\hbar \hat{\mathcal{P}}_{\pm}\,,\qquad
[\hat{\mathcal{D}},\hat{\mathcal{X}}_\pm]=-\frac{i\hbar}{2}\hat{\mathcal{X}}_\pm\,,\qquad 
[\hat{\mathcal{L}}_0,\hat{\mathcal{X}}_\pm]=\pm \frac{\hbar}{2}\hat{\mathcal{X}}_\pm \,, &\\&
[\hat{\mathcal{K}},\hat{\mathcal{P}}_\pm]=i\hbar\hat{\mathcal{X}}_{\pm}\,,
\qquad
[\hat{\mathcal{D}},\hat{\mathcal{P}}_\pm]=\frac{i\hbar}{2}\hat{\mathcal{P}}_\pm\,,\qquad 
[\hat{\mathcal{L}}_0,\hat{\mathcal{P}}_\pm]=\pm\frac{\hbar}{2}\hat{\mathcal{P}}_\pm\,,&\\&
[\hat{\mathcal{T}}_\pm,\hat{\mathcal{X}}_\mp]=-2i\hbar\hat{\mathcal{P}}_\pm\,,\qquad
[\hat{\mathcal{F}}_\pm,\hat{\mathcal{P}}_\mp]=2i\hbar\hat{\mathcal{X}}_\pm\,,&\\&
[\hat{\mathcal{L}}_\pm,\hat{\mathcal{X}}_{\mp}]=-i\hbar\hat{\mathcal{X}}_{\pm}\,,\qquad
[\hat{\mathcal{L}}_\pm,\hat{\mathcal{P}}_{\mp}]=i\hbar\hat{\mathcal{P}}_{\pm}\,.\label{Sch4}
&
\end{eqnarray}
On the other hand, from the non-relativistic limit applied 
to the Dirac equation
(\ref{Dirac}), 
one can read two superchargers 
\begin{eqnarray}
&\label{Pi12}
\hat{\Pi}_{1}=\frac{1}{\sqrt{2m}}\hat{\pi}\,,\qquad 
\hat{\Pi}_{2}=\frac{1}{\sqrt{2m}}i\sigma_3\hat{\pi}\,.
&
\end{eqnarray}
As the non-relativistic super-extended  Hamiltonian 
$\hat{\mathcal{H}}$ is invariant under the transformation
$\sigma_1(\hat{\mathcal{H}})\sigma_1$, one can  
 identify two more  
supercharges,
\begin{eqnarray}
&\label{Pi34}
\hat{\Pi}_3=\sigma_1(\hat{\Pi}_1)\sigma_1=\frac{1}{\sqrt{2m}}\left(\begin{array}{cc}
0 & \hat{p}_+\\
\hat{p}_- & 0 
\end{array}\right)\,,\qquad
\hat{\Pi}_4=\sigma_1(\hat{\Pi}_2)\sigma_1=-i\sigma_3\hat{\Pi}_3\,.
&
\end{eqnarray} 

For a future application  of the conformal bridge transformation, it is convenient  to 
introduce the following 
non-Hermitian combinations of the operators
 (\ref{Pi12}) and (\ref{Pi34}), 
\begin{eqnarray}
&\label{SuperPi4}
\hat{\Pi}_\pm^{(1)}=\frac{1}{2}(\hat{\Pi}_1\pm i\hat{\Pi}_2)=\frac{1}{\sqrt{8m}}\hat{p}_\pm\sigma_\mp\,,\qquad 
\hat{\Pi}_\pm^{(2)}=\frac{1}{2}(\hat{\Pi}_3\pm i\hat{\Pi}_4)=\frac{1}{\sqrt{8m}}\hat{p}_\pm\sigma_\pm\,.
&
\end{eqnarray}

Then, the  inclusion 
of these four odd  true
 integrals together with the four dynamical 
generators 
\begin{eqnarray}
&
\hat{\xi}_\pm^{(1)}=\sqrt{\frac{m}{8}}\hat{x}_\pm\sigma_\mp\,,\qquad
\hat{\xi}_\pm^{(2)}=\sqrt{\frac{m}{8}}\hat{x}_\pm\sigma_\pm\,,
&
\end{eqnarray}
and the 
order zero in momenta 
 true odd  integrals
$
\hat{\Sigma}_\pm=\sqrt{\frac{m}{8}}\sigma_\pm\
$ 
expands  the bosonic algebra  to its 
$\mathcal{N}=4$ supersymmetric 
extension. 
The remaining non-vanishing superalgebraic relations are  
\begin{eqnarray}
&\label{SUSYalge1}
[\hat{\mathcal{H}},\hat{\xi}_\pm^{(j)}]=-\hbar \hat{\Pi}_\pm^{(j)}\,,\qquad 
[\hat{\mathcal{K}},\hat{\Pi}_\pm^{(j)}]=\hbar \hat{\xi}_\pm^{(j)}\,,
\qquad
[\hat{\mathcal{L}}_0,\hat{\Pi}_\pm^{(2)}]=\frac{\hbar}{2} \hat{\Pi}_\pm^{(2)}
&\\&
[\hat{\mathcal{D}},\hat{\xi}_\pm^{(j)}]=-i\frac{\hbar}{2} \hat{\xi}_\pm^{(j)}\,,\qquad 
[\hat{\mathcal{D}},\hat{\Pi}_\pm^{(j)}]=i\frac{\hbar}{2} \hat{\Pi}_\pm^{(j)}\,,
\qquad 
[\hat{\mathcal{L}}_0,\hat{\xi}_\pm^{(2)}]=\frac{\hbar}{2} \hat{\xi}_\pm^{(2)}\,,
&\\&
[\hat{\mathcal{T}}_\pm,\hat{\xi}_\mp^{(1)}]=-i\hbar\hat{\Pi}_\pm^{(2)}\,,\qquad 
[\hat{\mathcal{T}}_\pm,\hat{\xi}_\mp^{(2)}]=-i\hbar\hat{\Pi}_\pm^{(1)}\,,
\qquad 
[\hat{\mathcal{R}},\hat{\Pi}_\pm^{(j)}]=\hbar \hat{\Pi}_\pm^{(j)}\,,
&\\&
[\hat{\mathcal{F}}_\pm,\hat{\Pi}_\mp^{(1)}]=i\hbar\hat{\xi}_\pm^{(2)}\,,\qquad 
[\hat{\mathcal{F}}_\pm,\hat{\Pi}_\mp^{(2)}]=i\hbar\hat{\xi}_\pm^{(1)}\,, \qquad 
[\hat{\mathcal{R}},\hat{\xi}_\pm^{(j)}]=\hbar \hat{\xi}_\pm^{(j)}\,,
&\\&
[\hat{\mathcal{L}}_\pm,\hat{\xi}_\mp^{(1)}]=-i\hbar\hat{\xi}_\pm^{(2)}\,,\qquad 
[\hat{\mathcal{L}}_\pm,\hat{\xi}_\mp^{(2)}]=-i\hbar\hat{\xi}_\pm^{(1)}\,,&\\&
[\hat{\mathcal{L}}_\pm,\hat{\Pi}_\mp^{(1)}]=i\hbar\hat{\Pi}_\pm^{(2)}\,,\qquad 
[\hat{\mathcal{L}}_\pm,\hat{\Pi}_\mp^{(2)}]=i\hbar\hat{\Pi}_\pm^{(1)}\,,
\qquad 
[\hat{\mathcal{R}},\hat{\Sigma}_\pm^{(j)}]=\hbar \hat{\Sigma}_\pm^{(j)}\,,
&\\&\label{PTsystemdef}
\{\hat{\Pi}_\pm^{(1)},\hat{\xi}_\mp^{(1)}\}=\hat{\mathcal{D}}\pm i(\hat{\mathcal{L}}_0+\frac{1}{2}\hat{\mathcal{R}})\,,\qquad 
\{\hat{\Pi}_\pm^{(2)},\hat{\xi}_\mp^{(2)}\}=\hat{\mathcal{D}}\pm i(\hat{\mathcal{L}}_0-\frac{3}{2}\hat{\mathcal{R}})\,,
&\\&
\{\hat{\zeta}_\pm^{(1)},\hat{\Pi}_\pm^{(2)}\}=
\{\hat{\zeta}_\pm^{(2)},\hat{\Pi}_\pm^{(1)}\}=\hat{\mathcal{L}}_\pm,
&\\&
\{\hat{\Pi}_{\pm}^{(1)},\hat{\Pi}_{\mp}^{(1)}\}=
\{\hat{\Pi}_{\pm}^{(2)},\hat{\Pi}_{\mp}^{(2)}\}=\hat{\mathcal{H}}\,,\qquad 
\{\hat{\xi}_{\pm}^{(1)},\hat{\xi}_{\mp}^{(1)}\}=
\{\hat{\xi}_{\pm}^{(2)},\hat{\xi}_{\mp}^{(2)}\}=\hat{\mathcal{K}}\,,
&\\&
\{\hat{\Pi}_{\pm}^{(1)},\hat{\Pi}_{\pm}^{(2)}\}=
\hat{\mathcal{T}}_\pm\,,\qquad 
\{\hat{\xi}_{\pm}^{(1)},\hat{\xi}_{\pm}^{(2)}\}=
\hat{\mathcal{F}}_\pm\,,\qquad 
\{\hat{\Sigma}_\pm,\hat{\Sigma}_\mp\}=\hat{\mathcal{M}}\,, &\\&
\{\hat{\xi}_\pm^{(1)},\hat{\Sigma}_\pm\}=\{\hat{\xi}_\pm^{(2)},\hat{\Sigma}_\mp\}=\frac{1}{2}\hat{\mathcal{X}}\,,\qquad
\{\hat{\Pi}_\pm^{(1)},\hat{\Sigma}_\pm\}=\{\hat{\Pi}_\pm^{(2)},\hat{\Sigma}_\mp\}=\frac{1}{2}\hat{\mathcal{P}}\,.
&\label{SUSYalgef}
\end{eqnarray}
From here  we note that the operators
$(\hat{\mathcal{H}},\hat{\mathcal{K}}, \hat{\mathcal{D}},\hat{\mathcal{L}}_0,\hat{\mathcal{L}}_\pm,\hat{\mathcal{T}}_\pm,\hat{\mathcal{F}}_\pm, 
\hat{\mathcal{R}} ,\hat{\Pi}_\pm^{(a)},\hat{\xi}_\pm^{(a)})$ with 
$a=1,2$ satisfy the orthosymplectic $D(1,2)\cong \mathfrak{osp}(2,4)$ superalgebra \cite{Dictionary}. 
On the other hand, 
the generators
$(\hat{\mathcal{X}}_\pm , \hat{\mathcal{P}}_\pm,\hat{\Sigma}_\pm,\hat{\mathcal{M}})$ 
constitute an ideal sub-superalgebra, which in turn is a 
supersymmetric extension of the two-dimensional Heisenberg symmetry with 
two additional fermionic ladder operators. 
 In conclusion, 
the entire supersymmetry of the system is a semi-direct sum of these two
 superalgebras.

\subsection{$2D$ 
superextension  of the Swanson model and  the spin-$1/2$ Landau problem}
\label{SubsecLandau}

To construct the Hamiltonian of  a
 $\mathcal{PT}$-invariant   
 supersymmetric system, we
just take a look at
the anti-commutation relations (\ref{PTsystemdef}) and 
consider
the
following four $\mathcal{PT}$-invariant operators 
\begin{eqnarray}
&\label{Options4}
\hat{\mathscr{H}}_{\mathcal{PT}}^{(\pm,1)}=2\omega(i\hat{\mathcal{D}}
\mp(\hat{\mathcal{L}}_0+\frac{1}{2}\hat{\mathcal{R}}) )\,,\qquad
\hat{\mathscr{H}}_{\mathcal{PT}}^{(\pm,2)}=2\omega(i\hat{\mathcal{D}}
\pm(\hat{\mathcal{L}}_0 - \frac{3}{2}\hat{\mathcal{R}}) )\,.\qquad
&
\end{eqnarray}
They satisfy relations
\begin{eqnarray}
&\label{2DSwanPTop}
\hat{\mathscr{H}}_{\mathcal{PT}}^{(\pm,1)}=-(\hat{\mathscr{H}}_{\mathcal{PT}}^{(\mp,1)})^{\dagger}\,,\quad 
\hat{\mathscr{H}}_{\mathcal{PT}}^{(\pm,2)}=-(\hat{\mathscr{H}}_{\mathcal{PT}}^{(\mp,2)})^{\dagger}\,,&\\&
\label{2DSwanPTop2}
\sigma_1(\hat{\mathscr{H}}_{\mathcal{PT}}^{(\pm,1)})\sigma_1=\hat{\mathscr{H}}_{\mathcal{PT}}^{(\pm,2)}\,,\qquad
&
\end{eqnarray}
and
it is enough to identify 
one of them as the $\mathcal{PT}$-invariant Hamiltonian.
We choose 
 \begin{equation}
 \label{def2DSwanPT}
 \hat{\mathscr{H}}_{\mathcal{PT}}^{(+,1)}:=
 \hat{\mathscr{H}}_{\mathcal{PT}}\,,
 \end{equation}
 and 
 relations (\ref{2DSwanPTop}), (\ref{2DSwanPTop2})  can be used then to obtain 
 the information for  the
 alternative models. 

As the 
CBT generator and  
the metric operator we take 
\be
\label{CBT2x2twodim}
\hat{\mathscr{S}}=\hat{\mathfrak{S}}\I_{2\cross 2}\,,\qquad
\hat{\mathscr{S}}^{-1}=\hat{\mathfrak{S}}^{-1}\I_{2\cross 2}\,,\qquad
\hat{\Theta}=(\hat{\mathscr{S}})^{2}\,.
\ee
Operator
 (\ref{def2DSwanPT}) 
is mapped  to 
 the Hermitian supersymmetric Hamiltonian 
\begin{eqnarray}\label{HLPTS}
&\hat{\mathcal{H}}_{L}:=
\hat{\mathscr{S}}(\hat{\mathscr{H}}_{\mathcal{PT}})\hat{\mathscr{S}}^{-1}=
\hat{H}_L\I_{2\cross2}-\hbar\omega\sigma_3\,,&
\end{eqnarray}
with
\begin{eqnarray}
&\label{HbosonicHarmonic}
\hat{H}_L=\hat{H}_{\text{osc}}-\omega\hat{p}_\varphi=
\hbar\omega(2\hat{b}_2^+\hat{b}_2^-+1)\,,
&
\end{eqnarray}
see Eqs. (\ref{NotationF}) and (\ref{b+b-}).
To interpret this system, 
we identify  
$\omega$ with the Larmor frequency
$\omega=\frac{eB}{2mc}$, where 
$q=-e$ is the electron charge and $B$ is a 
homogeneous magnetic field. In this case, 
  the operator $\hat{\mathcal{H}}_{L}$ takes the form of the Hamiltonian 
of a charged spin-1/2 non-relativistic  particle 
with  $g=2$  Land\'e factor    
subjected to
a homogeneous magnetic field, 
which is given by 
\begin{eqnarray}
&\label{SUSYLandauProb}
\hat{\mathcal{H}}_L=\frac{1}{2m}\left(\hat{\vp}-\frac{q}{c}\vA\right)^{2}-\hat{\vmu}\cdot\vB\,,\qquad 
\hat{\vmu}=-\frac{g\mu_{B}}{\hbar}\hat{\vS}\,,&\\&
A_{l}=\frac{B}{2}\epsilon_{lk}x_k\,,\quad
\epsilon_{lk}\partial_{l}A_k=-B\,,&
\end{eqnarray} 
where 
$\mu_{B}=\frac{e\hbar}{2mc}$ is the Bohr magneton, 
and $\hat{\vS}_{j}=\frac{\hbar}{2}\sigma_{j}$ is the  vector spin  operator. 
If we choose any  
other option
 in (\ref{Options4}) as the Hamiltonian of
the initial 
$\mathcal{PT}$-invariant model, we 
get 
a similar system  after the CBT
but with a change in the sign of the electric charge or in the direction of
the magnetic field, or with both changes simultaneously. 

On
 the other hand, we note that the  eigenstates  of (\ref{def2DSwanPT}) are given by
 the spinors
  \begin{eqnarray}
&\label{PTLandauState}
\Phi_{n_1,n_2}^{(+)}=\left(\begin{array}{c}
\phi_{n_1,n_2}\\
0
\end{array}\right)\,,
\quad 
\Phi_{n_1,n_2}^{(-)}=\left(\begin{array}{c}
0\\
\phi_{n_1,n_2}
\end{array}\right), 
\qquad n_1,n_2=0,1,\ldots\,,
&\\&\label{phin1n2}
\phi_{n_1,n_2}(x_1,x_2)=\frac{1}{\sqrt{2\pi n_1! n_2!}}\left(\frac{m\omega}{2\hbar}\right)^{\frac{n_1+n_2}{2}} z^{n_1}(z^*)^{n_2}\,,
\qquad z=x_1+ix_2\,,
&
\end{eqnarray} 
where $\phi_{n_1,n_2}(x_1,x_2)$ are the formal common eigenstates of the two-dimensional
bosonic operators $2i\hat{D}$ and $\hat{p}_\varphi$,
 that are mapped into the spin zero Landau problem 
eigenstates \cite{ERIHO}.
The basic properties that these functions  satisfy 
and which are necessary  to  compute the action of any operator  on the spinor states (\ref{PTLandauState})
are summarized  in Appendix \ref{AppPropPhi}. 
In particular, from here we deduce the eigenvalue equations 
\begin{eqnarray}\label{En2}
&
\hat{\mathscr{H}}_\mathcal{PT}\Phi_{n_1,n_2}^{(\pm)}=\hbar\omega(2n_2+1\mp 1)\Phi_{n_1,n_2}^{(\pm)}\,, \qquad
\hat{\mathcal{L}}_0 \Phi_{n_1,n_2}^{(\pm)}=\frac{\hbar}{2}(n_1-n_2\pm \frac{1}{2})\Phi_{n_1,n_2}^{(\pm)}\,.
\quad &
\end{eqnarray}
The spectrum of the system  
depends only  on 
the quantum number $n_2$, and therefore is infinite degenerate in each of its energy levels, including 
the ground state with zero energy. 
Additionally, there is an obvious supersymmetry here since the states 
$\Phi_{n_1,n_2+1}^{(+)}$ and 
$\Phi_{n_1,n_2}^{(-)}$ have  the same energy eigenvalue. 
 Accordingly,  
 the application of the CBT operator 
$\hat{\mathscr{S}}$ to
(\ref{PTLandauState})
yields the eigenstates of $\hat{\mathcal{H}}_L$ with the same energy eigenvalues, see 
Appendix \ref{AppPropPhi}.}

All the true
 and dynamical symmetries of the system (\ref{def2DSwanPT}) 
 can be identified by using the superalgebraic relations  
 (\ref{Simetryal1})--(\ref{Final?}), (\ref{Hei2})--(\ref{Sch4}) and 
(\ref{SUSYalge1})--(\ref{SUSYalgef}). Therefore, the application of the CBT to
 them 
will give
us the supersymmetry algebra  of the spin-1/2 Landau problem (\ref{SUSYLandauProb}). 
For the sake of simplicity, 
we restrict ourselves to look for the true integrals of motion related with 
the degeneracy of the spectrum. 
To do that, we take into
 account that the Hamiltonian 
is a complex linear combination of the generator of dilatations
and the total angular momentum operator 
$2\hat{\mathcal{L}}_0+\hat{\mathcal{R}}$
of the spin-$1/2$ particle.
So, 
$\hat{\mathcal{L}}_0$ and $\hat{\mathcal{R}}=\frac{\hbar}{2}\sigma_3$ commute with
$\hat{\mathscr{H}}_{\mathcal{PT}}$. In addition to them,  
one finds that
 \begin{eqnarray}
 &\label{BosonicLandau}
 \hat{\mathscr{B}}_+=\sqrt{\frac{\omega}{2\hbar m}}\hat{\mathcal{X}}_+\,,\quad 
 \hat{\mathscr{B}}_-= \frac{1}{\sqrt{2\hbar m\omega}}i \hat{\mathcal{P}}_-\,,\qquad
 \hat{\mathscr{G}}_+=\frac{\omega}{2\hbar}\hat{\mathcal{F}}_+\,,\quad 
 \hat{\mathscr{G}}_-= -\frac{1}{2\hbar \omega} \hat{\mathcal{T}}_-
 \quad
 &
  \end{eqnarray}
are the true bosonic integrals of motion of $\hat{\mathscr{H}}_\mathcal{PT}$. 
Under CBT, they are mapped 
into the bosonic operators
 \begin{eqnarray} 
 &
 \hat{\mathscr{S}}^{-1}(\hat{\mathscr{B}}_\pm, \hat{\mathscr{G}}_\pm)\hat{\mathscr{S}}^{-1}=
 (\hat{b}_1^{\pm}\I_{N\cross N},\frac{1}{2}(\hat{b}_1^{\pm})^2\I_{N\cross N})\,,
  \end{eqnarray}
where 
$\hat{b}_1^\pm$ 
correspond to  
 complex linear  combinations of the integrals that 
 represent the center of the  circular  orbit of the particle
  in the classical Landau problem. 
  They are
   the first order
  ladder operators for the quantum number
  $n_1$ in the  eigenfunctions, 
  not appearing
in the 
energy spectrum  (\ref{En2})  
   \cite{ERIHO}. 
  
In order to identify the supercharges of the system, we just take a look at
  the first relations in   (\ref{SUSYalge1})
 and
 (\ref{PTsystemdef}), which 
with the identification
\begin{eqnarray}
&\label{FermionicLandau}
\hat{Q}_-=2i
\hat{\Pi}_+^{(1)}
=\frac{i}{\sqrt{2 
m}}\hat{p}_+\sigma_-\,,\qquad
\hat{Q}_+=
2\omega\hat{\xi}_-^{(1)}=
\sqrt{\frac{m \omega^{2}}{2}}
\hat{x}_-\sigma_+
& 
\end{eqnarray} 
can be rewritten
 as the $\mathcal{N}=2$ Poincar\'e superalgebra,
\begin{eqnarray}
[\hat{\mathscr{H}}_{\mathcal{PT}},\hat{Q}_\pm]=0\,,\qquad
\{\hat{Q}_-,\hat{Q}_+\}=2\hat{\mathscr{H}}_{\mathcal{PT}}\,,\qquad 
\{\hat{Q}_\pm,\hat{Q}_\pm\}=0\,.\quad 
\end{eqnarray}
The
odd operators $\hat{Q}_\pm$ are recognised as the supercharges of the system, 
which
are 
responsible for the 
supersymmetry mentioned above, 
\begin{eqnarray}
&
\hat{Q}_\mp\Phi_{n_1,n_2}^{(\pm)}=\sqrt{\hbar\omega(n_2+\frac{1\mp1}{2})}\Phi_{n_1,n_2\mp 1}^{(\mp)}\,,\quad 
\hat{Q}_\pm \Phi_{n_1,n_2}^{(\pm)}=0\,.\quad
&
\end{eqnarray}
Then
  under the 
  CBT 
   we have 
\begin{eqnarray}
&
\hat{\mathscr{S}}(\hat{Q}_\pm)\hat{\mathscr{S}}^{\pm}=\hat{\mathcal{Q}}_\pm\,,\qquad 
\hat{\mathcal{Q}}_\pm=\sqrt{\hbar\omega}\hat{b}_2^\pm \sigma_\pm\,,
&
\end{eqnarray}
where  the operators
 $\sqrt{\hbar\omega}\hat{b}_2^\pm$ can be identified as the 
complex    combinations  of the components of the vector operator $\hat{\vp}-\frac{q}{c}\vA$
in the corresponding Landau problem
\cite{ERIHO}.

Next, let us take a look at what happens 
with  the other fermionic generators of the free particle system. 
 First, due to the presence of $\hat{\mathcal{R}}$ in the Hamiltonian $\hat{\mathscr{H}}_\mathcal{PT}$, 
the
operators $\hat{\Sigma}_\pm$ cannot be true integrals of the system. 
On the other hand, 
the application of 
the Hermitian conjugation and the unitary  transformation generated by $\sigma_1$ 
to supercharges (\ref{FermionicLandau}) produces
\begin{eqnarray}
&\label{SuperModels1}
-(\hat{Q}_-)^\dagger=2i\hat{\Pi}_-^{(1)}\,,\qquad
-(\hat{Q}_+)^\dagger=2\omega \hat{\xi}_+^{(1)}\,,\qquad
&\\&
\label{SuperModels2}
\sigma_1(\hat{Q}_-)\sigma_1=2i\hat{\Pi}_+^{(2)}\,,\qquad
\sigma_1(\hat{Q}_+)\sigma_1=2\omega \hat{\xi}_-^{(2)}\,,
&\\&
\label{SuperModels3}
-(\sigma_1(\hat{Q}_-)\sigma_1)^{\dagger}=2i\hat{\Pi}_-^{(2)}\,,\qquad
-(\sigma_1(\hat{Q}_+)\sigma_1)^\dagger=2\omega \hat{\xi}_+^{(2)}\,.
&
\end{eqnarray}
In this way, in Eqs.  (\ref{SuperModels1}), 
(\ref{SuperModels2}) and (\ref{SuperModels3})
there appear the remaining odd generators 
that come from the free particle system, 
and they are the supercharges of the Hamiltonians 
$\hat{\mathscr{H}}_\mathcal{PT}^{(-,1)}$,
$\hat{\mathscr{H}}_\mathcal{PT}^{(+,2)}$
and 
$\hat{\mathscr{H}}_\mathcal{PT}^{(-,2)}$, respectively, 
see (\ref{PTsystemdef}). Using these arguments 
and the commutation relations introduced in the previous subsection, 
it is easy to see that none  of these odd operators 
commutes  with $\hat{\mathscr{H}}_\mathcal{PT}$.
So,  all  they are the dynamical integrals.

In conclusion, operators (\ref{BosonicLandau}) and (\ref{FermionicLandau}) 
are all the true symmetry generators that the system possesses, 
and they encode all the special  properties
of the spectrum. The remaining integrals  are 
of  the
dynamical nature, and they work as the
ladder operators~\footnote{ 
This can be shown from the superalgebra presented in subsection \ref{SubsecNonRecLim}, where 
one can see that each operator is an eigenstate of $\hat{\mathcal{D}}$, $\hat{\mathcal{L}}_0$ and  $\hat{\mathcal{R}}$
in the sense of the  adjoint action 
$[\hat{O},\hat{A}_i]=\lambda_i \hat{A}_i$.
}.
Unlike 
the free particle case, we 
have here the $\mathcal{N}=2$ instead of 
$\mathcal{N}=4$
supersymmetric extension as we  explained above.

%%%%%%%%%%
%%%%%%%%%%%
%%%%%%%%%%%%
%%%%%%%%%%%%%
%%%%%%%%%%%%%%
%%%%%%%%%%%%%
%%%%%%%%%%%%
%%%%%%%%%%%

\subsection{Inclusion of an Aharonov-Bohm flux} 
\label{SubSecAharonov-Bohm}

Let us consider now a 
spin zero 
 electrically charged particle coupled to a
curl-free  in the exterior of a solenoid
vector potential 
\begin{eqnarray}
&
{\vA}^{\Delta
 \Phi}= \frac{\Delta
  \Phi}{2\pi \rho^2}(-x_2,x_1,0)\,,\qquad
\vnabla\cross\vA^{\Delta 
\Phi}=0\,. 
\qquad 
\rho^2=x_1^2+x_2^2\,,&
\end{eqnarray}
where $\Delta 
\Phi$ is the total magnetic flux inside the solenoid. 
The Hamiltonian operator of such a system has the form 
\begin{eqnarray}
\label{AhraFunc}
&
\hat{H}^{(\alpha)}=\frac{1}{2m}(\hat{p}_i+\frac{e}{c}A_{i}^{\Delta 
\Phi})^{2}=
\hat{H}+\frac{\hbar\alpha}{2m\rho^2}(\hbar\alpha+2\hat{p}_\varphi)\,,\qquad 
\alpha=\frac{e\Delta
 \Phi}{2\pi c \hbar}\,. 
&
\end{eqnarray}
As we will see below, the model is solvable for arbitrary values of the dimensionless
 parameter $\alpha$.
However, it allows for
different interpretations depending on the 
values  
of this number.  
The case $\alpha\in \Z$ appears in the study of super-conductivity
and quantum Hall effect, and corresponds to a quantized magnetic flux \cite{Hall}. 
On the other hand, 
Hamiltonian (\ref{AhraFunc}) with arbitrary values of the parameter $\alpha$
describes the free dynamics in relative coordinate of the system 
of two identical 
anyons \cite{LeiMyr,MacWil,CalAny1}.

By 
solving the corresponding Schr\"odinger equation, we find the eigenstates 
of the system, as well as its energy eigenvalues 
\begin{eqnarray}
&\label{FreePart}
\psi_{\kappa,\ell}^{(\alpha)}=\sqrt{\frac{\kappa}{2\pi}}J_{|\ell+\alpha|}(\kappa \rho)e^{i\ell \varphi}\,,\qquad
E_{\kappa}=\frac{\hbar^2\kappa^2}{2m}\,,\qquad
\kappa>0\,,\quad \ell=0,\pm1,\ldots
&
\end{eqnarray}
where $\varphi$ is the angular coordinate.
 
Such a model 
can be 
obtained from the usual two-dimensional free particle via 
the 
unitary singular at the origin transformation 
\cite{Hall}
\begin{eqnarray}
&
\hat{U}_\alpha \hat{H}\hat{U}_\alpha^{\dagger}=\hat{H}^{(\alpha)}\,,\qquad 
\hat{U}_\alpha \hat{p}_\varphi\hat{U}_\alpha^{\dagger}=\hat{p}_\varphi+\hbar\alpha\,:=\,\,
\hat{p}_\varphi^{(\alpha)}\,,\quad
&
\end{eqnarray}
where
\be 
\label{U}
\hat{U}_\alpha=e^{-i\alpha \varphi}\,,\qquad
\hat{U}_\alpha^{\dagger}=e^{i\alpha \varphi}\,.
\ee
Unlike $\hat{H}$ and $\hat{p}_\varphi$, operators
$\hat{K}$ and $\hat{D}$ are invariant under this transformation, 
and the system  $\hat{H}_\alpha$  still maintains the  $\mathfrak{so}(2,1)$ conformal invariance. 

Operator $\hat{p}_\varphi^{(\alpha)}$ can be understood as the total angular momentum 
of the system, with eigenvalues $\hbar(\ell+\alpha)$. 
In the same vein, 
the first order operators $\hat{p}_\pm$
and $\hat{x}_\pm$ transform as
\begin{eqnarray}
&\label{MomentaOpmodi}
\hat{U}_\alpha(\hat{p}_\pm)\hat{U}_\alpha^\dagger=\hat{p}_\pm^{(\alpha)}=-i\hbar e^{\pm i\varphi}[\frac{\partial}{\partial \rho}
\mp\frac{1}{\hbar\rho}\hat{p}_\varphi^{(\alpha)}]\,,\qquad
\hat{U}_\alpha(\hat{x}_\pm)\hat{U}_\alpha^\dagger=\hat{x}_\pm\,.
&
\end{eqnarray}
However,
 $\hat{p}_\pm^{(\alpha)}$ are physical operators 
in the Hilbert space generated by 
the states  (\ref{FreePart}) only when $\alpha\in \Z$, see Appendix
\ref{AppPonPhiPsi}
\footnote{System (\ref{AhraFunc}) 
and the two-dimensional free particle are unitary equivalent 
iff $\alpha\in \Z$. 
At $\alpha\neq \Z$, operator $\hat{U}_\alpha$ acting on the free particle eigenstate 
$J_{|\ell|}(\kappa \rho)e^{i \ell\varphi}$ produces the multi-valued function 
$J_{|\ell|}(\kappa \rho)e^{i (\ell+\alpha)\varphi}$ which is different from 
 (\ref{FreePart}).
}.
Despite these  peculiarities associated with values of the parameter 
$\alpha$ related  to  the mapping of symmetry generators, 
we will maintain it as an arbitrary real number for a while. 
Then, by  considering the action of the matrix
extension of this unitary transformation 
$\hat{U}_\alpha\rightarrow \hat{\mathcal{U}}_\alpha=\hat{U}_\alpha\I_{2\cross 2}$ on
$\mathcal{PT}$-symmetric Hamiltonian operator (\ref{def2DSwanPT}), we get 
  \begin{eqnarray}
&\label{2dPt1AB}
\hat{\mathscr{H}}_{\mathcal{PT}}^{(\alpha)}=\hat{\mathcal{U}}_\alpha (\hat{\mathscr{H}}_{\mathcal{PT}})
\hat{\mathcal{U}}_\alpha ^{\dagger}
=
2\omega(i\hat{\mathcal{D}}\mp(\hat{\mathcal{L}}_0+\frac{1}{2}\hat{\mathcal{R}}+\frac{\hbar}{2}\alpha) )\,,
&
\end{eqnarray}
which in effect, is just the $\hat{\mathscr{H}}_{\mathcal{PT}}$ 
shifted for
the constant $\hbar\omega \alpha$. 
To apply the conformal bridge transformation to the
system (\ref{2dPt1AB}), 
we use the modified operator
 \begin{eqnarray}
 & 
 \hat{\mathscr{S}}_\alpha= \hat{\mathcal{U}}_\alpha (\hat{\mathscr{S}})\hat{\mathcal{U}}_\alpha^\dagger=
 \hat{\mathfrak{S}}_\alpha\I_{2\cross 2}\,,\qquad
 \hat{\mathfrak{S}}_\alpha=e^{\frac{\pi}{4\omega\hbar}(\hat{H}^{(\alpha)}-\omega^2\hat{K})}\,,
\end{eqnarray}  
and its 
inverse. 
As a result  we obtain 
the Hermitian Hamiltonian operator 
\begin{eqnarray}
&\label{SpinSolLan}
\hat{\mathscr{S}}_\alpha(\hat{\mathscr{H}}_{\mathcal{PT}}^{(\alpha)})\hat{\mathscr{S}}_\alpha^{-1}= 
\hat{H}_L^{(\alpha)}\I_{2\cross 2}-\hbar\omega\sigma_3 \, : = \,\hat{\mathcal{H}}_{L}^{(\alpha)}\,,&
\end{eqnarray} 
where 
\begin{eqnarray}
& \label{HLalpha}
\hat{H}_L^{(\alpha)}= \hat{U}_\alpha \hat{H}_L  \hat{U}_\alpha^{\dagger} =
\hat{H}+\frac{\hbar\alpha}{2m\rho^2}(\hbar\alpha+2\hat{p}_\varphi)+\frac{m\omega^2}{2} \rho^2-
\omega\hat{p}_\varphi-\hbar \omega \alpha\,.
&\end{eqnarray}
Here, (\ref{HLalpha})  has the form 
of the Hamiltonian of a scalar 
charged particle 
in external magnetic field described by
 the total vector potential $
{\vA_T}={\vA}+{\vA}^{\Delta 
\Phi}\,.
$ 
So, (\ref{SpinSolLan}) is 
 a generalization of  
Hamiltonian (\ref{HLPTS}) for the case in the presence of 
the Aharonov-Bohm flux,
\begin{eqnarray}
&
\hat{\mathcal{H}}_{L}^{(\alpha)}=\frac{1}{2m}\left(\hat{\vp}+\frac{e}{c} {\vA_T}\right)^2-\hat{\vmu}\cdot\vB\,.
&
\end{eqnarray}

To construct the eigenstates, we
look 
for the structure of the spinors in polar coordinates. 
For this, we introduce the functions 
\begin{eqnarray}
&\label{Jordan4}
\phi_{n_\rho,\ell}^{(\alpha)}(r,\varphi)=\mathcal{N}_{n_\rho,\ell}^{(\alpha)}\rho^{2n_\rho+|\ell+\alpha|}e^{i\ell\varphi} \,,
\quad 
\mathcal{N}_{n_\rho,\ell}^{(\alpha)}=
(-1)^{n_\rho}\frac{\left(\frac{m\omega}{2\hbar}\right)^{n_\rho+\frac{|\ell+\alpha|}{2}}}
{\sqrt{\pi n_\rho ! \Gamma(n_\rho+|\ell+\alpha|+1)}}\,, &\\& n_\rho=0,1,\ldots\,,\quad
\ell=0,\pm 1,\ldots\,.\qquad&
\end{eqnarray}
They satisfy the relations
\begin{eqnarray}
&
2i\omega\hat{D}\phi_{n_\rho,\ell} ^{(\alpha)}(r,\varphi) =
\hbar\omega(2n_\rho+|\ell+\alpha|+1)\phi_{n_\rho,\ell} ^{(\alpha)}(r,\varphi)\,, 
&\\&\label{JordanAgain}
\hat{p}_\varphi\phi_{n_\rho,\ell}^{(\alpha)}(r,\varphi)=\hbar\ell\phi_{n_\rho,\ell}^{(\alpha)}(r,\varphi)\,, \qquad
\frac{1}{\hbar\omega}\hat{H}^{(\alpha)}\phi_{n_\rho,\ell}^{(\alpha)}=\sqrt{n_\rho(n_\rho+|\ell+\alpha|)}\phi_{n_\rho-1,\ell}^{(\alpha)}\,,
&
\end{eqnarray}
and the last equation in (\ref{JordanAgain})
helps us to show that the corresponding CBT produces 
\begin{eqnarray}
&
e^{\frac{\pi}{4\omega\hbar}(\hat{H}_\alpha-\omega^2\hat{K})}\phi_{n_\rho,\ell}^{(\alpha)}(\rho,\varphi)=
\psi_{n_\rho,\ell}^{(\alpha)}(\rho,\varphi)\,,
&\\&\label{LLLStates} 
\psi_{n_\rho,\ell}^{(\alpha)}(\rho,\varphi)=\left(\frac{m\omega}{\hbar^2}\right)^\frac{1}{2}
\sqrt{\frac{n_\rho!}{\pi\Gamma(n_\rho+|\ell+\alpha|+1)}}\zeta^{|\ell|}
L_{n_\rho}^{(|\ell+\alpha|)}(\zeta^2)e^{-\frac{\zeta^2}{2}}e^{i\ell\varphi}\,,\qquad 
\zeta=\sqrt{\frac{m\omega}{\hbar}}\rho\,,\quad
&
\end{eqnarray}
where $L_n^{(\beta)}(\eta)$ are the generalized Laguerre polynomials. 
These functions are the eigenstates of the operator (\ref{HLalpha}), 
\begin{eqnarray}
\hat{H}_{L}^{(\alpha)}\psi_{n_\rho,\ell}^{(\alpha)}(\rho,\varphi)=
\hbar\omega(2n_\rho+(1-\text{sign}(\ell+\alpha))|\ell+\alpha|+1)
\psi_{n_\rho,\ell}^{(\alpha)}(\rho,\varphi)\,. 
\end{eqnarray}
Functions (\ref{Jordan4}) are useful to construct the spinor states of 
the
 $\mathcal{PT}$-invariant 
supersymmetric system 
(\ref{2dPt1AB}),  and to identify the real spectrum of the model,  
\begin{eqnarray}
&
\label{eigen2dPolar}
\Phi_{n_\rho,\ell}^{(+,\alpha)}(\rho,\varphi)=\left(\begin{array}{c}
\phi_{n_\rho,\ell}^{(\alpha)}(\rho,\varphi)\\
0
\end{array}\right)\,,\qquad 
\Phi_{n_\rho,\ell}^{(-,\alpha)}(\rho,\varphi)=\left(\begin{array}{c}
0\\
\phi_{n_\rho,\ell}^{(\alpha)}(\rho,\varphi)\\
\end{array}\right)\,,
&\\&
\hat{\mathscr{H}}_{\mathcal{PT}}^{(\alpha)}\Phi_{n_\rho,\ell}^{(\pm,\alpha)}(\rho,\varphi)=
\hbar\omega(2n_\rho+(1-\text{sign}(\ell+\alpha))|\ell+\alpha|+1\mp 1)
\Phi_{n_\rho,\ell}^{(\pm,\alpha)}(\rho,\varphi)\,.&
\end{eqnarray}

All the true  symmetry and supersymmetry generators of this 
 $\mathcal{PT}$-invariant
 model with $\alpha\in\Z$
 can be obtained via the unitary transformation generated by 
$\hat{\mathcal{U}}_\alpha$ and applied to 
those of  the system discussed in 
 Sec. \ref{SubsecLandau}. 
For $\alpha\neq\Z$,
 the action of the momenta operators (\ref{MomentaOpmodi})
on states (\ref{Jordan4}) is not well-defined as it is shown in Appendix \ref{AppPonPhiPsi},
and consequently linear symmetry generators are not allowed.
If there exist supersymmetry  generators for the non-trivial cases $\alpha\notin\Z$, they 
have to be obtained using a different approach.  
We comment on this point in the last section.
%%%%
%%%%%%%%%
%%%%%%%%%%%%%%%%%%%%

\subsection{Supersymmetric ERIHO system}
\label{SubSecERIHO}
We start this subsection
 by analyzing  the $\mathcal{PT}$-invariant
Hamiltonian operator  
\begin{eqnarray}
&
\label{PT-her}
\hat{\mathscr{H}}_\gamma=2\omega\left(i\hat{\mathcal{D}}+\gamma\left[\hat{\mathcal{L}}_0 +\frac{1}{2} \hat{\mathcal{R}}\right]\right) -\hbar\omega(\gamma+1)\,,
&
\end{eqnarray}
where 
$\gamma$ is an arbitrary  real parameter. 
In	 the case $\gamma=0$ we obtain two copies of the 
dilatation generator,  while
the $\mathcal{PT}$-symmetric model (\ref{def2DSwanPT}) is recovered
in the case $\gamma=-1$. 
The application of the CBT yields the model described by the Hamiltonian
\begin{eqnarray}
&
\label{SUSYERIHO}
\hat{\mathcal{H}}_\gamma=\hat{\mathfrak{S}}(\hat{\mathscr{H}}_{\gamma})\hat{\mathfrak{S}}^{-1}= 
\hat{H}_{\text{osc}}+2\gamma\omega\left[\hat{\mathcal{L}}_0 +\frac{1}{2} \hat{\mathcal{R}}\right] -\hbar\omega(\gamma+1)\,,
&
\end{eqnarray}
which in matrix form is given by 
\begin{eqnarray}
\hat{\mathcal{H}}_\gamma=\left(\begin{array}{cc}
\hat{H}_{\text{osc}}+\gamma\omega\hat{p}_\varphi-\hbar\omega & 0\\
0 & \hat{H}_{\text{osc}}+\gamma\omega\hat{p}_\varphi-\hbar\omega(2\gamma+1)
\end{array}\right)\,.
\end{eqnarray}
This Hermitian Hamiltonian operator 
is a supersymmetric extension of the ERIHO system \cite{ERIHO},
whose spin zero version is described by the Hamiltonian operator
\begin{eqnarray}
\label{ERIHO}
\hat{H}_\gamma=\hat{H}_{\text{osc}}+\gamma\omega \hat{p}_\varphi=
\hbar\omega[(1+\gamma)\hat{b}_1^{+}\hat{b}_1^{-}+(1-\gamma)\hat{b}_2^{+}\hat{b}_2^{-}+1]\,.
\end{eqnarray}
It can be related with
 the bosonic Landau problem
 subjected to an additional harmonic force, 
or with the harmonic oscillator system in a  rigidly rotating reference frame, see \cite{ERIHO}. 
As in the case $\gamma=-1$, 
the eigenstates of the system (\ref{PT-her})
are also given by Eq. (\ref{eigen2dL}), but now,
they satisfy   
the eigenvalue equations
 \begin{eqnarray}
 &
\label{Hphi=EPhi}
\hat{\mathscr{H}}_\gamma \Phi_{n_1,n_2}^{(+)}=
E_{n_1,n_2}^{(\gamma)}\Phi_{n_1,n_2}^{(+)}\,,\qquad 
\hat{\mathscr{H}}_\gamma \Phi_{n_1+1,n_2-1}^{(-)}=
E_{n_1,n_2}^{(\gamma)}\Phi_{n_1+1,n_2-1}^{(-)}\,,&\\&
\hat{\mathscr{H}}_\gamma \Phi_{0,0}^{(-)}=E_{0,0}^{*(\gamma)}\Phi_{0,0}^{(-)}\,,\quad
E_{0,0}^{*(\gamma)}=-2\gamma\hbar\omega\,,\quad
E_{n_1,n_2}^{(\gamma)}=\hbar\omega(n_1+n_2+\gamma(n_1-n_2))\,.\quad
&
\end{eqnarray}
Note that  when $|\gamma|<1$, the spectrum 
is bounded from  below  and in this case we can always redefine the Hamiltonian operator 
in order to have a ground energy level equal to zero 
when
$0<\gamma\leq 1$. 
Otherwise, the spectrum is not bounded from below, and even if
there is a non-degenerated eigenstate with zero energy eigenvalue, 
 this spinor cannot 
be identified as the ground state of the system (the model is ill-defined
within
the usual framework of quantum mechanics). On the other hand, 
from equation  
  (\ref{Hphi=EPhi})
we note the existence of  supersymmetry for arbitrary values of $\gamma$.  
The fermionic  (odd) integrals 
 responsible for
  this phenomenon
correspond to
\begin{eqnarray}
\label{intI}
&
\hat{I}_+=\frac{i}{2\hbar}\hat{x}_+\hat{p}_+\sigma_-\,,\quad
\hat{I}_-=\frac{i}{2\hbar}\hat{x}_-\hat{p}_-\sigma_+\,,\quad 
\hat{\mathfrak{S}}(\hat{I}_\pm)\hat{\mathfrak{S}}^{-1}=
\hat{b}_1^\pm \hat{b}_2^\mp \sigma_\mp :=\hat{\mathcal{I}}_\pm
\,,\quad 
&
\end{eqnarray} 
which simultaneously commute with $\hat{\mathcal{D}}$ and $2\hat{\mathcal{L}}_0+\hat{\mathcal{R}}$, and
 their  action on states  (\ref{eigen2dL}) yields   
\begin{eqnarray}
&
\hat{I}_+\Phi_{n_1,n_2}^{(+)}=2\sqrt{n_2(n_1+1)}\Phi_{n_1+1,n_2-1}^{(-)}\,,\quad
\hat{I}_-\Phi_{n_1,n_2}^{(-)}=2\sqrt{n_1(n_2+1)}\Phi_{n_1-1,n_2+1}^{(+)}\,,\quad&\\&
\hat{I}_\pm\Phi_{n_1,n_2}^{(\mp)}=0\,.  &
\end{eqnarray}  
This supersymmetry does not enter in contradiction with the fact that we can have 
negative eigenvalues since the anti-commutator  of supercharges 
$\hat{I}_\pm$ 
produces not the Hamiltonian (\ref{PT-her}) but the 
quadratic polynomial in it and in the even integral 
$2\hat{\mathcal{L}}_0+\hat{\mathcal{R}}$,
\begin{eqnarray}
&
\{\hat{I}_-,\hat{I}_+\}=
\frac{1}{\hbar^2}(2i\hat{\mathcal{D}}-(2\hat{\mathcal{L}}_0+\hat{\mathcal{R}}))
(2i\hat{\mathcal{D}}+(2\hat{\mathcal{L}}_0+\hat{\mathcal{R}}))\,,
&\\&
\{\hat{I}_+,\hat{I}_-\}\Phi_{n_1,n_2}^{(\pm)}=(2n_2+1\mp 1)(2n_1+1\pm 1)\Phi_{n_1,n_2}^{(\pm)}\,. 
&
\end{eqnarray}
Note that in the case  $\gamma=-1$ studied already in Sec. \ref{SubsecLandau},
the $\hat{I}_\pm$ are not independent integrals since
\begin{eqnarray}
&
\hat{I}_+=\frac{1}{\sqrt{2\hbar\omega}}\hat{\mathscr{B}}_+\hat{Q}_-\,,\qquad 
\hat{I}_-=\frac{1}{\sqrt{2\hbar\omega}}\hat{\mathscr{B}}_-\hat{Q}_+\,.
&
\end{eqnarray}

In  Ref.
\cite{ERIHO}, it  was shown that the bosonic model 
(\ref{ERIHO}) possesses hidden symmetry integrals of motion
 only when $\gamma$
is a rational number.
So,  the obvious question here is whether our extended model has additional 
fermionic integrals for particular values of this parameter.   
In the following, we will 
consider the case $0<|\gamma|\leq 1$, and after that 
we study the case   $|\gamma|>1$.  In both cases 
we obtain 
higher order fermionic operators whose action on the spinors
(\ref{PTLandauState}) is reconstructed  with the help of the basic formulas from 
 Appendix \ref{AppPropPhi}. 
\begin{itemize}
\item
\textit{
{\bf i)} The case $\gamma=\frac{s_2-s_1}{s_2+s_1}$, with
 $s_1,s_2=1,2,\ldots,$ and $|\gamma|\leq 1$.}
\end{itemize}
When we  choose $\gamma$ in this way, one can show that the operators
 \begin{eqnarray}
 &\label{IntL}
 \hat{L}_{s_1,s_2}^{+}=(\sqrt{\frac{m\omega}{2\hbar}}\hat{x}_{+})^{s_1}(\frac{i}{\sqrt{2m\omega\hbar}}\hat{p}_{+})^{s_2}\I_{2\cross 2}\,,\quad 
 \hat{L}_{s_1,s_2}^{-}=(\frac{i}{\sqrt{2m\omega\hbar}}\hat{p}_{-})^{s_1}(\sqrt{\frac{m\omega}{2\hbar}}\hat{x}_{-})^{s_2}\I_{2\cross 2}\,,\qquad  
 & 
\end{eqnarray}  
commute with our Hamiltonian 
(\ref{PT-her}). 
They together with $\hat{\mathscr{H}}_\gamma$ and 
$\hat{\mathcal{L}}_{0}$ generate a nonlinear deformation of the 
$\mathfrak{su}(2)$ algebra in the general case \cite{ERIHO},
and by CBT generators (\ref{CBT2x2twodim}) are transformed
into
\begin{eqnarray}
&
\hat{\mathscr{S}}(\hat{L}_{s_1,s_2}^{\pm}) \hat{\mathscr{S}}^{-1}=\hat{\mathcal{L}}_{s_1,s_2}^\pm\,,\qquad
\hat{\mathcal{L}}_{s_1,s_2}^\pm=(\hat{b}_1^\pm)^{s_1}(\hat{b}_2^\mp)^{s_2}\I_{2\cross 2}\,.
&
\end{eqnarray}

The action 
of (\ref{IntL}) on the eigenstates of the system is
given by
\begin{eqnarray}
&
 \hat{L}_{s_1,s_2}^{+}\Phi_{n_1,n_2}^{(\pm)}=\sqrt{\frac{n_2!\Gamma(n_1+s_1+1)}{n_1!\Gamma(n_2-s_2+1)}}
\Phi_{n_1+s_1,n_2-s_2}^{(\pm)}\,,&\\&
 \hat{L}_{s_1,s_2}^{-}\Phi_{n_1,n_2}^{(\pm)}=\sqrt{\frac{n_1!\Gamma(n_2+s_2+1)}{n_2!\Gamma(n_1-s_1+1)}}
\Phi_{n_1-s_1,n_2+s_2}^{(\pm)}\,.
&
\end{eqnarray}
To find additional supercharges for the choice 
 of $\gamma$ under consideration, 
we compute the commutator of integrals (\ref{IntL})
 with the odd integrals
 $\hat{I}_\pm$. As a result one gets 
\begin{eqnarray}
&
[ \hat{L}_{s_1,s_2}^{\pm},\hat{I}_\mp]=\frac{i(s_2-s_1)}{\hbar}(2\hat{D}\mp i\frac{s_2+s_1}{s_2-s_1}\hat{p}_\varphi+i\hbar 
+\frac{2i\hbar s_1 s_2}{s_2-s_1})\hat{Q}_{s_1,s_2}^\pm\,,
&\\&
[ \hat{L}_{s_1,s_2}^{\pm},\hat{I}_\pm]=0\,,
&
\end{eqnarray} 
where 
\begin{eqnarray}
&\label{Qs1s2Qs1s2}
\hat{Q}_{s_1,s_2}^+=\frac{1}{2}(\sqrt{\frac{m\omega}{2\hbar}}\hat{x}_+)^{s_1-1}(\frac{i}{\sqrt{2m\omega\hbar}}\hat{p}_+)^{s_2-1}\sigma_+\,,
\quad
&\\&
\hat{Q}_{s_1,s_2}^-=\frac{1}{2}(\frac{i}{\sqrt{2m\omega\hbar}}\hat{p}_-)^{s_1-1}(\sqrt{\frac{m\omega}{2\hbar}}\hat{x}_-)^{s_2-1}\sigma_-\,,
&
\end{eqnarray} 
which under CBT are mapped 
as 
$\hat{\mathscr{S}}(\hat{Q}_{s_1,s_2}^{\pm}) \hat{\mathscr{S}}^{-1}=\hat{\mathcal{Q}}_{s_1,s_2}^\pm$, 
$\hat{\mathcal{Q}}_{s_1,s_2}^\pm
=\frac{1}{2}(\hat{b}_{1}^{\pm})^{s_1-1}
(\hat{b}_{2}^{\mp})^{s_2-1}\sigma_\pm
$.
Note that the existence of these operators assumes
 that $s_1,s_2\geq 1$. 
The action of the
resulting operators 
 on the eigenstates produces 
\begin{eqnarray}
&
\label{QpmonPhi}
\hat{Q}_{s_1,s_2}^-\Phi_{n_1,n_2}^{(+)}=\sqrt{\frac{n_1!\Gamma(n_2+s_2)}{n_2!\Gamma(n_1-s_1+2)}}
\Phi_{n_1-s_1+1,n_2+s_2-1}^{(-)}\,,
\qquad 
\hat{Q}_{s_1,s_2}^\pm\Phi_{n_1,n_2}^{(\pm)}=0\,,
\qquad 
&\\&
\hat{Q}_{s_1,s_2}^+\Phi_{n_1,n_2}^{(-)}=\sqrt{\frac{n_2!\Gamma(n_1+s_1)}{n_1!\Gamma(n_2-s_2+2)}}
\Phi_{n_1+s_1-1,n_2-s_2+1}^{(+)}\,. 
&
\end{eqnarray}
We do not explicitly calculate the anti-commutator between these 
two supercharges. Instead, let us comment on some properties of the 
resulting operator. First, due to the structure of (\ref{Qs1s2Qs1s2}), 
one sees that the anti-commutator 
has to be
 a polynomial operator in terms of $\hat{x}_\pm$, $\hat{p}_\pm$ 
and $\sigma_3$ only. 
Besides, from  relations 
(\ref{QpmonPhi}) we can calculate the action of this anti-commutator on the eigenstates, 
which gives us
\begin{eqnarray}
&\label{antiQonPhi1}
\{\hat{Q}_{s_1,s_2}^+,\hat{Q}_{s_1,s_2}^-\}\Phi_{n_1,n_2}^{(+)}=\frac{\Gamma(n_1+1)}{\Gamma(n_1-s_1+2)}\frac{\Gamma(n_2+s_2)}{\Gamma(n_2+1)}
\Phi_{n_1,n_2}^{(+)}\,, &\\&
\label{antiQonPhi2}
\{\hat{Q}_{s_1,s_2}^+,\hat{Q}_{s_1,s_2}^-\}
\Phi_{n_1,n_2}^{(-)}=
\frac{\Gamma(n_2+1)}{\Gamma(n_2-s_2+2)}\frac{\Gamma(n_1+s_1)}{\Gamma(n_1+1)}
\Phi_{n_1,n_2}^{(-)}\,.
&
\end{eqnarray}
These equations show 
that the operator 
$\{\hat{Q}_{s_1,s_2}^+,\hat{Q}_{s_1,s_2}^-\}=\hat{O}$
is bounded from below
and annihilates the eigenstates of the form  $\Phi_{n_1,n_2}^{(+)}$ with $n_1-s_1+2\leq 0$
and $\Phi_{n_1,n_2}^{(-)}$ with  $n_2-s_2+2\leq 0$. 
Finally, as a consequence of the relations 
\begin{eqnarray}
&
[\hat{\mathcal{L}}_0,\hat{Q}_{s_1,s_2}^\pm]=\pm\frac{\hbar}{2}(s_1+s_2-1)\hat{Q}_{s_1,s_2}^\pm\,,\qquad
[2i\hat{\mathcal{D}},\hat{Q}_{s_1,s_2}^\pm]=\pm\hbar(s_1-s_2)\hat{Q}_{s_1,s_2}^\pm\,,
&\\&
[\hat{\mathcal{R}},\hat{Q}_{s_1,s_2}^\pm]=\pm\hbar\hat{Q}_{s_1,s_2}^\pm\,,&
\end{eqnarray}
one can see that $[\hat{\mathcal{L}}_0,\hat{O}]=[2i\hat{\mathcal{D}},\hat{O}]=[\hat{\mathcal{R}},\hat{O}]=0$.
These relations imply that $\hat{O}=\hat{O}(\hat{\mathcal{D}},\hat{\mathcal{R}},\hat{\mathcal{L}}_0)$
since this structure is the only possible option that satisfies all the already mentioned
 properties  by using the operators
 $\hat{x}_\pm$, $\hat{p}_\pm$ 
and $\sigma_3$ as the building blocks.

On the other hand, 
\begin{eqnarray}
&
[ \hat{L}_{s_1,s_2}^{\pm},\hat{Q}_{s_1,s_2}^\pm]=0\,,\quad 
[ \hat{L}_{s_1,s_2}^{\pm},\hat{Q}_{s_1,s_2}^\mp]=P(\hat{\mathscr{H}_\gamma},\hat{\mathcal{L}}_0,\hat{\mathcal{R}})
\hat{I}_{\pm}\,,\quad
\{\hat{Q}_{s_1,s_2}^\pm,\hat{I}_{\pm}\}=2\hat{L}_{s_1,s_2}^{\pm}\,,&\qquad
\end{eqnarray}
where $P(\hat{\mathscr{H}_\gamma},\hat{\mathcal{L}}_0,\hat{\mathcal{R}})$ is another 
model dependent
polynomial function.  
Therefore, we learn that the set of generators 
$(\hat{\mathscr{H}_\gamma},\hat{\mathcal{L}}_0,\hat{\mathcal{R}}, \hat{L}_{s_1,s_2}^\pm, 
\hat{Q}_{s_1,s_2}^\pm,\hat{I}_{\pm})$ satisfy a nonlinear superalgebra
that completely  describes 
 the degeneracy of the spectrum.   

\begin{itemize}
\item
\textit{ {\bf ii)}
The case $\gamma=\frac{s_2+s_1}{s_2-s_1}$ with
$s_1,s_2=1,2,\ldots,$ and $|\gamma|\geq 1$
}.
\end{itemize}
Here one can show that the operators 
\begin{eqnarray}
&
\hat{J}_{s_1,s_2}^+=(\frac{m\omega}{2\hbar})^{\frac{s_1+s_2}{2}}(\hat{x}_+)^{s_1}(\hat{x}_-)^{s_2}\,,\qquad 
\hat{J}_{s_1,s_2}^-=(\frac{-1}{m\omega\hbar})^{\frac{s_1+s_2}{2}}(\hat{p}_-)^{s_1}(\hat{p}_+)^{s_2}\,,
\label{Js1Js2}
&\\&
\hat{\mathscr{S}}(\hat{J}_{s_1,s_2}^{\pm}) \hat{\mathscr{S}}^{-1}=\hat{\mathcal{J}}_{s_1,s_2}^\pm\,,\qquad
\hat{\mathcal{J}}_{s_1,s_2}^\pm=(\hat{b}_1^\pm)^{s_1}(\hat{b}_2^\pm)^{s_2}\I_{2\cross 2}\,, 
&
\end{eqnarray}
commute with the Hamiltonian (\ref{PT-her}) for this choice
of $\gamma$. 
These integrals are 
responsible for
the infinite degeneracy 
of each energy level since $
\hat{J}_{s_1,s_2}^{+}$ cannot annihilate any eigenstate,  
\begin{eqnarray}
&
\hat{J}_{s_1,s_2}^{+}\Phi_{n_1,n_2}^{(\pm)}=\sqrt{\frac{\Gamma(n_1+s_1+1)\Gamma(n_1+s_2+1)}{n_1!n_2!}}
\Phi_{n_1+s_1,n_2+s_1}^{(\pm)}\,, &\\&
\hat{J}_{s_1,s_2}^{-}\Phi_{n_1,n_2}^{(\pm)}=\sqrt{\frac{n_1!n_2!}{\Gamma(n_1-s_1+1)\Gamma(n_1-s_2+1)}}
\Phi_{n_1-s_1,n_2-s_1}^{(\pm)}\,.
&
\end{eqnarray}

Then 
we 
compute 
the  commutators  of 
 these integrals with
the odd integrals $\hat{I}_\pm$, which as a result  give
us 
four higher order fermionic integrals, 
\begin{eqnarray}
&
[\hat{J}_{s_1,s_2}^\pm,\hat{I}_\pm]=\mp s_2\hat{W}_{s_1,s_2}^{\pm}\,,\qquad 
[\hat{J}_{s_1,s_2}^\pm ,\hat{I}_\mp ]=\mp s_1\hat{T}_{s_1,s_2}^\pm\,,\qquad &\\& 
\hat{W}_{s_1,s_2}^{+}=(\frac{m\omega}{2\hbar})^{\frac{s_1+s_2}{2}}(\hat{x}_+)^{s_1+1}(\hat{x}_-)^{s_2-1}\frac{\sigma_-}{2}\,,\quad 
\hat{T}_{s_1,s_2}^{+}=(\frac{m\omega}{2\hbar})^{\frac{s_1+s_2}{2}}(\hat{x}_+)^{s_1-1}(\hat{x}_-)^{s_2+1}\frac{\sigma_+}{2}\,,\qquad 
&\\&
\hat{W}_{s_1,s_2}^{-}=(\frac{-1}{m\omega\hbar})^{\frac{s_1+s_2}{2}}(\hat{p}_-)^{s_1+1}(\hat{p}_+)^{s_2-1}\frac{\sigma_+}{2}\,,\quad
\hat{T}_{s_1,s_2}^{-}=(\frac{-1}{m\omega\hbar})^{\frac{s_1+s_2}{2}}(\hat{p}_-)^{s_1-1}(\hat{p}_+)^{s_2+1}\frac{\sigma_-}{2}\,.\qquad
&
\end{eqnarray}
Acting on them, the CBT produces 
\begin{eqnarray}
&
\hat{\mathscr{S}}(\hat{W}_{s_1,s_2}^{\pm},\hat{T}_{s_1,s_2}^{\pm}) \hat{\mathscr{S}}^{-1}=(\hat{\mathcal{W}}_{s_1,s_2}^\pm\,,
\hat{\mathcal{T}}_{s_1,s_2}^\pm)\,,\qquad &\\&
\hat{\mathcal{W}}_{s_1,s_2}^\pm=\frac{1}{2}(\hat{b}_1^\pm)^{s_1+1}(\hat{b}_2^\pm)^{s_2-1}\sigma_\mp\,,\qquad 
\hat{\mathcal{T}}_{s_1,s_2}^\pm=\frac{1}{2}(\hat{b}_1^\pm)^{s_1-1}(\hat{b}_2^\pm)^{s_2+1}\sigma_\pm\,. 
&
\end{eqnarray}

Note that the existence of  the operators 
$\hat{W}_{s_1,s_2}^{\pm}$ ($\hat{T}_{s_1,s_2}^{\pm}$)
assumes that $s_2>0$ ($s_1>0$). 
We also note  here that
the case  $s_2=0$ ($s_1=0$) corresponds 
to the case $\gamma=-1$ ($\gamma=1$),  which was 
analyzed 
in subsection \ref{SubsecLandau}.

The action of these operators produces
\begin{eqnarray}
&
\hat{W}_{s_1,s_2}^{+}\Phi_{n_1,n_2}^{(+)}=\sqrt{\frac{\Gamma(n_1+s_1+2)\Gamma(n_2+s_2)}{n_1!n_2!}}\Phi_{n_1+s_1+1,n_2+s_2-1}^{(-)}\,,
&\\&
\hat{W}_{s_1,s_2}^{-}\Phi_{n_1,n_2}^{(-)}=\sqrt{\frac{n_1!n_2!}{\Gamma(n_1-s_1)\Gamma(n_2-s_2+2)}}\Phi_{n_1-s_1-1,n_2-s_2+1}^{(+)}\,,
&\\&
\hat{T}_{s_1,s_2}^{+}\Phi_{n_1,n_2}^{(-)}=\sqrt{\frac{\Gamma(n_1+s_1)\Gamma(n_2+s_2+2)}{n_1!n_2!}}\Phi_{n_1+s_1-1,n_2+s_2+1}^{(+)}\,,
&\\&
\hat{T}_{s_1,s_2}^{-}\Phi_{n_1,n_2}^{(+)}=\sqrt{\frac{n_1!n_2!}{\Gamma(n_1-s_1+2)\Gamma(n_2-s_2)}}\Phi_{n_1-s_1+1,n_2-s_2-1}^{(-)}\,,
&\\&
\hat{W}_{s_1,s_2}^{\pm}\Phi_{n_1,n_2}^{(\mp)}=
\hat{T}_{s_1,s_2}^{\pm}\Phi_{n_1,n_2}^{(\pm)}=0\,.
&
\end{eqnarray}
We do not compute the anti-commutators, which are essentially 
 nonlinear. Instead of that, we consider the action of the
anti-commutators on
the eigenstates, 
\begin{eqnarray}
&
\{\hat{W}_{s_1,s_2}^{+},\hat{W}_{s_1,s_2}^{-}\}\Phi_{n_1,n_2}^{(+)}=\frac{\Gamma(n_1+s_1+2)\Gamma(n_2+s_2)}{n_1!n_2!}\Phi_{n_1,n_2}^{(+)}\,,
&\\&
\{\hat{W}_{s_1,s_2}^{+},\hat{W}_{s_1,s_2}^{-}\}\Phi_{n_1,n_2}^{(-)}=\frac{n_1!n_2!}{\Gamma(n_1-s_1+2)\Gamma(n_2-s_2)}\Phi_{n_1,n_2}^{(-)}\,,
&\\&
\{\hat{T}_{s_1,s_2}^{+},\hat{T}_{s_1,s_2}^{-}\}\Phi_{n_1,n_2}^{(-)}=\frac{\Gamma(n_1+s_1)\Gamma(n_2+s_2+2)}{n_1!n_2!}\Phi_{n_1,n_2}^{(-)}\,,
&\\&
\{\hat{T}_{s_1,s_2}^{+},\hat{T}_{s_1,s_2}^{-}\}\Phi_{n_1,n_2}^{(+)}=\frac{n_1!n_2!}{\Gamma(n_1-s_1)\Gamma(n_2-s_2+2)}\Phi_{n_1,n_2}^{(+)}\,.
&
\end{eqnarray} 
One can note
 that even if the eigenstate has a negative energy level, 
there is no negative eigenvalues in these four equations. 
This implies that the explicit form of the commutators have
to
be, similarly to 
 the case 
$|\gamma|<1$,
 a polynomial of the bosonic operators that 
 does not have negative eigenvalues.

Some of the superalgebraic relations  that 
describe
the system are
\begin{eqnarray}
&
\{\hat{W}_{s_1,s_2}^{\pm},\hat{T}_{s_1,s_2}^{\pm}\}=(\hat{J}_{s_1,s_2}^\pm)^{2}\,,\qquad
\{\hat{W}_{s_1,s_2}^{\pm},\hat{T}_{s_1,s_2}^{\mp}\}=0\,,\qquad \qquad
&\\&
\{\hat{W}_{s_1,s_2}^{\pm},\hat{I}_{\mp}\}=P_1(\hat{\mathcal{D}},\hat{\mathcal{L}}_0,\hat{\mathcal{R}})\hat{J}_{s_1,s_2}^\pm\,,\qquad
\{\hat{T}_{s_1,s_2}^{\pm},\hat{I}_{\pm}\}=P_2(\hat{\mathcal{D}},\hat{\mathcal{L}}_0,\hat{\mathcal{R}})\hat{J}_{s_1,s_2}^\pm\,,\qquad
&\\&
\label{alge1}
[\hat{J}_{s_1,s_2}^\mp,\hat{W}_{s_1,s_2}^\pm]=P_3(\hat{\mathcal{D}},\hat{\mathcal{L}}_0,\hat{\mathcal{R}})\hat{I}_{\pm}\,,\qquad 
[\hat{J}_{s_1,s_2}^\mp,\hat{T}_{s_1,s_2}^\pm]=P_4(\hat{\mathcal{D}},\hat{\mathcal{L}}_0,\hat{\mathcal{R}})\hat{I}_{\pm}\,,\qquad 
&\\&
\label{alge2}
[2i\hat{\mathcal{D}},\hat{W}_{s_1,s_2}^\pm]=\pm \hbar(s_1+s_2)\hat{W}_{s_1,s_2}^\pm\,,\qquad 
[2i\hat{\mathcal{D}},\hat{T}_{s_1,s_2}^\pm]=\pm \hbar(s_1+s_2)\hat{T}_{s_1,s_2}^\pm\,, &\\&
\label{alge3}
[\hat{\mathcal{L}}_0,\hat{W}_{s_1,s_2}^\pm]=\pm \frac{\hbar}{2}(s_1-s_2+1)\hat{W}_{s_1,s_2}^\pm\,,\qquad 
[\hat{\mathcal{L}}_0,\hat{W}_{s_1,s_2}^\pm]=\pm \frac{\hbar}{2}(s_1-s_2-1)\hat{T}_{s_1,s_2}^\pm\,, \quad&\\&
\label{alge4}
[\hat{\mathcal{R}},\hat{W}_{s_1,s_2}^\pm]=\mp \hbar\hat{W}_{s_1,s_2}^\pm\,,\qquad 
[\hat{\mathcal{R}},\hat{W}_{s_1,s_2}^\pm]=\pm \hbar \hat{T}_{s_1,s_2}^\pm\,,
&
\end{eqnarray}
where $P_k(\hat{\mathcal{D}},\hat{\mathcal{L}}_0,\hat{\mathcal{R}})$,
$k=1,2,3,4$, are polynomial functions. 
From  relations (\ref{alge2})--(\ref{alge4}) one can 
verify that  the 
fermionic operators 
$\hat{W}_{s_1,s_2}^{\pm}$  and $\hat{T}_{s_1,s_2}^{\pm}$
are integrals 
of motion of the system.

In conclusion of this section we note that all the 2D super-extended  
 $\mathcal{PT}$-symmetric systems
and their Hermitian analogs produced by applying the CBT 
that we have considered in this section 
are superintegrable. In particular  case of the supersymmetrically 
extended Landau problem each of its two superpartners has four true integrals 
of motion. In the Hermitian version, these are the corresponding Hamiltonian,
angular momentum, and the non-commuting  operators being quantum analogs of 
the classical coordinates  of the center of the circular orbit described by  $b^+_1$ and $b^-_1$.
These four integrals, however,   are functionally dependent, 
while  the Hamiltonian, angular momentum, and  one of the two last integrals or their 
linear combination form a set  of three independent
integrals of motion. In the cases {\bf i)}  and {\bf ii)}   of the super-extended ERIHO systems
we have a similar picture with  components  of the diagonal 
bosonic integrals (\ref{IntL}) and (\ref{Js1Js2}), respectively,
 which play a role analogous to $b^+_1$ and $b^-_1$.

\section{Discussion and Outlook}
\label{SecDisc}

In the light of the presented results, 
we 
 list  here  some problems 
related to them that   
may be interesting for 
a future research.

\textbf{1.} According   to the Riemann hypothesis,  nontrivial zeroes of the Riemann  zeta function
lie on the critical line and have 
the form $\frac{1}{2} +  iE_n$
 with $E_n \in \R $.  The Hilbert-P\'olya conjecture argues  
  that the real values $E_n$  correspond  to 
  eigenvalues of a self-adjoint  operator.
 An important  result  that gave credibility to
this conjecture was obtained by  Selberg \cite{Selberg},
who established a  bridge  (a kind of duality) between the quantum 
spectrum of the Laplacian on compact Riemann surfaces of 
constant negative curvature and the length spectrum of their 
prime geodesics. Selberg trace formula, which establishes that link, 
strongly resembles Riemann explicit formula. 
The next step was put forward by Berry
who proposed the Quantum Chaos conjecture, according
to which the Riemann zeros are the 
spectrum of a Hamiltonian 
obtained by quantization of a classical chaotic
Hamiltonian, whose periodic orbits are labelled by the
prime numbers \cite{Berry}. 
In 1999 Berry and Keating \cite{BerKea,BerKea+},   and Connes \cite{Connes} 
suggested that a spectral realization of
the Riemann zeros could be achieved by quantizing the 1D classical Hamiltonian
$H = xp$, which generates dilatations in  the phase space ($x$, $p$).  
Remarkably,  generator of dilatations also plays a key role
in constructing  the compact hyperbolic Riemann surfaces
of a constant curvature and geodesics on them  from the hyperbolic Lobachevsky plane 
\cite{McKean,Marklof}.  

Different quatization schemes  of the classical system $H = xp$
were considered in the literature in the context of the Hilbert-P\'olya conjecture, see, e.g. 
 \cite{Sierra}
and further references therein.
Probably, Bender and collaborators came closest to the solution of the problem in \cite{BenderRieman,BenderRieman2}
by employing  the theory and ideas 
of the $\mathcal{PT}$-symmetry.
In their construction, 
the non-Hermitian operator 
(natural units are assumed here)
\begin{eqnarray}
\label{HBen}
&
\hat{\H}=\hat{\Delta}^{-1}\{\hat{x},\hat{p}\}\hat{\Delta} \,,\qquad
\hat{\Delta}=1-e^{-i\hat{p}}\,,\qquad x\in \R^+\,,
&
\end{eqnarray}
is employed
  as  an alternative quantization prescription  of the Berry-Keating (B-K) Hamiltonian. 
Its eigenfunctions $\psi_n(x)$, $x\in (0,\infty)$, subject to the boundary condition
$\psi_n(x)=0$,  are characterized by eigenvalues $\{ E_n \}$ such that
$\{ \frac{1}{2}(1-iE_n)\}$ are the nontrivial zeros of the Riemann zeta function
in agreement with the B-K conjecture.
On the other hand, the trivial zeros $z=-2n$,  $n=1,2, \ldots $,  of the Riemann zeta function 
are associated with real eigenvalues of the operator $i \hat{\H} $ characterized by the unbroken 
$\mathcal{PT}$-symmetry. 
This can be seen  
from the explicit form of the eigenstates they use, which 
are of the form 
\begin{eqnarray}
&\label{ReiamnnEigen}
\psi_z(x)=\hat{\Delta}^{-1}x^{-z}=-\zeta(z,x+1)\,,\qquad 
\hat{\H}\psi_z(x)=i(2z-1)\psi_z(x)\,,
&
\end{eqnarray}
where $\zeta(z,a)$ is the Hurwitz function that reduces to the Riemann zeta function when 
$a=1$.  

Note that the operator (\ref{HBen}) is (up to a numerical factor and a multiplication by $-i$) 
our $\mathcal{PT}$-symmetric Hamiltonian (\ref{Swanson2iD}) 
subjected to a similarity transformation generated by $\hat{\Delta}$
and restricted to the half-axis.
To relate these ideas with our constructions of Sec. 2, 
we realize the $\mathfrak{so}(2,1)\cong\mathfrak{sl}(2,\R)$ symmetry generators 
as 
\begin{eqnarray}
&\label{so21gennu}
\hat{H}_\nu=\frac{1}{2}\hat{p}^2+\frac{\nu(\nu+1)}{2x^2}\,,\qquad 
\hat{D}=\frac{1}{4}\{\hat{x},\hat{p}\}\,,
\qquad 
\hat{K}=\frac{1}{2}\hat{x}^{2}\,,
&\\&
x\in\R^+\,,\qquad
\nu\geq -\frac{1}{2}\,.
&
\end{eqnarray}
In this way, the CBT
produces the operator \cite{Inzunza:2019sct,InzPly9}
\begin{eqnarray}
&\label{CBTAFF}
\hat{\mathfrak{S}}_{\nu}(2i\hat{D})\hat{\mathfrak{S}}_\nu^{-1}=\frac{1}{2}\left(\hat{p}^2+\frac{\nu(\nu+1)}{x^2}+x^2\right)
:=\hat{H}_\nu^{\text{osc}}\,,
\qquad
 \hat{\mathfrak{S}}_\nu= \exp(\frac{\pi}{4}(\hat{H}_\nu-\hat{K}))\,,
&
\end{eqnarray}
corresponding to the harmonically confined version of the one-dimensional conformal mechanics 
 model. As a consequence, one finds eigenvalues and  relations  between corresponding eigenstates,
 \begin{eqnarray}
 &
 2i\hat{D}\phi_{n,\nu}=E_{n,\nu}\phi_{n,\nu}\,\,\Rightarrow\,\, \hat{H}_{\nu}^{\text{osc}}\psi_{n,\nu}=
 E_{n,\nu}\psi_{n,\nu}\,,\quad
 E_{n,\nu}=2n+\nu+\frac{3}{2}\,,&\\&
 \phi_{n,\nu}=x^{\nu+1+2n}\,\,\Rightarrow \,\,
 \psi_{n,\nu}(x)\propto \hat{\mathfrak{S}}_\nu \phi_{n,\nu}\propto x^{\nu+1} L_{n}^{(\nu+\frac{1}{2})}(x^2)e^{-x^2/2}\,,\quad 
 n=0,1,\ldots\,.
& 
 \end{eqnarray}
 On the other hand, since 
 formally   
\begin{eqnarray}
&\label{654}
\hat{\mathfrak{S}}_\nu\hat{\Delta}(i\hat{\H})\hat{\Delta}^{-1}\hat{\mathfrak{S}}_\nu^{-1} =2\hat{H}_{\nu}^{\text{osc}}\,,&
\end{eqnarray} 
according to \cite{BenderRieman,BenderRieman2}  one gets  
 that 
$\hat{\Delta}^{-1}\hat{\mathfrak{S}}_\nu^{-1}\psi_{n,\nu}\propto
\hat{\Delta}^{-1}x^{1+\nu+2n}=-\zeta(-2n-\nu-1,x+1)$
 are the eigenstates of $i\hat{\H}$ with eigenvalues $2E_{n,\nu}$. Then 
the trivial zeros of the Riemann zeta function 
are obtained at $x=0$ when $\nu=-1$.  

In this way, the equation (\ref{654}) establishes a relationship between the study of 
the Riemann hypothesis and the model of the harmonically confined conformal mechanics, 
and we believe that this observation may  contribute somehow 
to this area of research. One may also wonder 
if something interesting might happen for other values of the parameter $\nu$
distinguishing distinct $\mathfrak{sl}(2,\R)$ representations. 
This, however requires further investigation.

\textbf{2.} The geometric background on which the dynamics takes place 
can affect the physical properties of the system. In particular, the presence of explicit and hidden symmetries essentially 
depends on it 
\cite{Carter,GRH,CarFKK,CarigliaRev,Frolov3}. 
In this context, one may wonder what forms the symmetries and supersymmetries 
of models like those discussed in the last section will take if we modify the 
space-time.
For example, one can study the case of conical geometry,
 which formally can be
 obtained from our Hermitian systems 
(\ref{HbosonicHarmonic}) and (\ref{ERIHO}) 
via the local canonical transformation 
\cite{Inzunza:2021vlb,InzPly7}
\begin{eqnarray}
\rho\rightarrow\alpha\rho\,,\qquad 
p_\rho\rightarrow\alpha^{-1}p_\rho\,,\qquad
\varphi\rightarrow\alpha^{-1}\varphi\,,\qquad
p_\varphi\rightarrow\alpha p_\varphi\,,
\end{eqnarray}
where $\alpha$ is a real parameter whose interpretation depends on its numerical value.
Under this transformation, the Euclidean metric in polar coordinates 
$ds^{2}=d\rho^2+\rho^{2} d\varphi^2$ takes the form  
\begin{eqnarray}
\label{conicalGeo}
ds_\alpha^{2}=\alpha^2d\rho^2+\rho^{2} d\varphi^2\,.
\end{eqnarray}
So, when $\alpha=\csc^2(\beta)>1$, one interprets $\beta$ as the angular aperture of a cone. 
On the other hand,
for
$0<\alpha<1$, 
 (\ref{conicalGeo}) can be understood 
as  the metric of 
a background with a
radial dislocation.
Under further
 identification 
$\alpha=(1-4\mu c^{-2}G)^{-1}$, one can say that (\ref{conicalGeo}) represents the spatial 
part of a cosmic string background with a positive ($\alpha>1$) 
or negative ($0<\alpha<1$) linear  mass density $\mu$. 
For the bosonic cases of the planar harmonic oscillator and the ERIHO system in this space, 
we showed in \cite{InzPly7} 
 that the list of  integrals of motion depends on the value of this
  parameter. Consequently, one can expect
that these
 geometrical properties have to control
 the appearance of fermionic integrals in the supersymmetric case.

\textbf{3.} The following natural step is to explore generalizations of the 
two-dimensional picture discussed 
in Sec. \ref{SecSUSYCBT2d} 
to higher dimensions. In this case one can 
start by looking the non-relativistic  limit of the 
 free Dirac and Klein-Gordon 
equations in  $(d+1)$ Minkowski  space. 
An obvious generalization of the corresponding
$\mathfrak{so}(2,1)$ conformal symmetry operators, on which 
 the CBT generators are based, will be  
\begin{eqnarray}
\hat{\mathcal{H}}=\hat{H}\I_{d'\cross d'}\,,\qquad
\hat{\mathcal{D}}=\hat{D}\I_{d'\cross d'}\,,\qquad
\hat{\mathcal{K}}=\hat{K}\I_{d'\cross d'}\,,
\end{eqnarray}
where $\hat{H}$, $\hat{D}$ and $\hat{K}$ are the $d$-dimensional generators (\ref{HKDind}),
and $d'$ is the dimension of representation of the corresponding 
Dirac matrices. Then the  interesting point 
 here is to find  the fermionic operators 
by using the Clifford algebra generators
as well as the momenta and position operators.
We  expect that 
the anti-commutators  between 
the
odd generators will be similar to those in  (\ref{PTsystemdef}), 
that 
could help to identify possible 
 candidates for a higher-dimensional 
$\mathcal{PT}$-invariant 
supersymmetric Hamiltonian operator. 
It would be interesting to identify the integrals of 
motion of the resulting model and their possible  
quantum mechanical interpretation  in
 the light of the conformal bridge transformation. 

\textbf{4.} Returning  
to the case of the Aharonov-Bohm
 flux addressed in  Sec. \ref{SubSecAharonov-Bohm}, 
we believe that the non-existence of well-defined momenta operators 
 for the non-trivial case $\alpha\notin \Z$ does not necessarily  imply 
 a lack of supersymmetry in our systems. 
 The principal indication in this direction 
 is the already known fact that the spinless 
  case possesses  hidden superconformal symmetry 
 for arbitrary values of $\alpha$ 
 \cite{CorPLyAB}.  That was  shown 
 by considering the reflection 
 operator $\hat{R}$, corresponding to 
 a rotation  for angle $\pi$,
 as the grading operator. 
 With respect to it, the well-defined in the  integer
 case $\alpha=n$ 
 operators  $\hat{p}_\pm^{(\alpha)}$
 are taken as the  supercharges that anticommute  for
 the Hamiltonian
 $\hat{H}^{(\alpha)}$. For $\alpha\not=n$,  
 the  von Neumann theory on self-adjoint  extensions 
 allowed there
 to identify 
  the nonlocal operators $\hat{p}_1 + i\epsilon \hat{R} \hat{p}_2$ and
 $-\epsilon \hat{p}_1 + i \hat{R} \hat{p}_2$   
 as  the well-defined supercharges of the system,
where
 $\epsilon=\pm$ depends periodically on the value range of  $\alpha\in\R$.
 Additionally, in the half-integer case the model has hidden 
 supersymmetries generated by higher order nonlocal 
operators. 
 On the other hand, it is also remarkable that
  if we restrict the particle to move on 
 a circle, 
the system  preserves the hidden  Poincar\'e supersymmetry for 
 $\alpha$ integer and half-integer as well \cite{CorPly}. Besides, 
 it was shown then in  \cite{JakNiePly,InzPly1}  that such 
 hidden Poincar\'e supersymmetry  
and hidden superconformal symmetry  in purely 
bosonic systems originate
from the corresponding 
super-exended systems with  fermion degrees of freedom by applying to them  a 
special  nonlocal  unitary transformation  followed by an appropriate  reduction procedure.
It was also observed  that exotic supersymmetry 
 appears in some self-isospectral 
\cite{JakNiePly+,NiePly}
and anyon  \cite{CFJP} systems  with fermion degrees of freedom  under the choice of  nonlocal grading operators. 
We  speculate then  that by taking into account 
the aspects of the theory of self-adjoint  extensions used in ref.  \cite{CorPLyAB},
 and allowing nonlocality  in grading operators, 1) our spin-1/2  
 $\mathcal{PT}$-invariant system (\ref{2dPt1AB})  can be 
generalized to admit supersymmetry for arbitrary values of the parameter $ \alpha $,
and 2) that the hidden superconformal symmetry 
in some  purely bosonic systems can be treated on the base of the
$\mathcal{PT}$-symmetry
and conformal bridge transformations.

\textbf{5.} Finally, it would be interesting
 to extend our  construction
 to the relativistic case. 
In this context, let us return to the problem discussed
 in Sec. \ref{SubsecLandau}, 
and consider the fermionic operators (\ref{FermionicLandau}) to construct the 
pseudo-Hermitian operator $\hat{Q}=ic\sqrt{m}\sigma_3(\hat{Q}_-+\hat{Q}_+)$. 
Under the conformal bridge transformation, this operator maps into 
\begin{eqnarray}
&\label{DiacarHamA}
\hat{\mathscr{S}}(\hat{Q}+
mc^2\sigma_3)\hat{\mathscr{S}}^{-1}=c\sigma_{j}(\hat{p}_{j}
-\frac{q}{c}A_{j})+mc^2\sigma_3\,,\qquad A_{j}=-\frac{cm\omega}{q}\epsilon_{jk}
\hat{x}_k\,,
&
\end{eqnarray} 
which corresponds to a Dirac Hamiltonian of the form (\ref{Dirac0}),  
coupled to a homogeneous magnetic field  with vector potential 
$A_i$.
We could follow the logic of this article and postulate 
$ \hat{Q} + mc^2 \sigma_3 $ as a first-order 
pseudo-Hermitian but not $\mathcal{PT}$-invariant Hamiltonian, 
even though this is not
a usual form of a Dirac Hamiltonian
since the CBT also works as a Dyson map for it.  
In order to have a more familiar structure, 
we can then consider another
kind of similarity transformations to relate this operator 
with a Dirac Hamiltonian plus a non-Hermitian interaction term. 
An example of that could
be 
\begin{eqnarray}
&\label{Transforamed}
e^{\frac{\mathcal{H}}{\omega\hbar}}(2^{-\frac{1}{2}}\hat{Q}+mc^2\sigma_3)e^{-\frac{\mathcal{H}}{\omega\hbar}}
=c\sigma_{j}\hat{p}_{j}+i\frac{m\omega c}{2}\hat{x}_-\sigma_+ + mc^2\sigma_3\,,
&
\end{eqnarray}
which now looks like 
a pseudo-Hermitian Dirac Hamiltonian with a non-Hermitian interaction term.  
This new system can then be related  to 
(\ref{DiacarHamA}) via the composition of the  
transformation inverse  to (\ref{Transforamed})
 and the CBT. Such an approach
 can be useful to generalize
 the ideas of Swanson and other authors on
  pseudo-Hermitian non-relativistic systems 
 for the case of Dirac-like  models.

\vskip0.3cm

\noindent {\large{\bf Acknowledgements} } 
\vskip0.1cm

The work was partially supported by the FONDECYT Project 1190842,
 FONDECYT Project 3220327, 
and the DICYT Project 042131P\_POSTDOC.

 \appendix
  \section{Equivalence through similarity transformations}
 \label{AppLabel}
Here   we explicitly construct the similarity transformations that relate the
$\mathcal{PT}$-symmetric operator $2i\Omega \hat{D}$ with the 
Swanson model (\ref{SwansonInSO1})
in dependence on the values of its parameters $\alpha$ and $\beta$.
The construction only involves algebraic arguments, 
and  so it is valid for any realization
of the conformal generators employed in the article. 
 
Based on  algebra (\ref{so21gen+}), we write down the  relations
\begin{eqnarray}
&
e^{a\hat{H}}\hat{D} 
e^{-a\hat{H}}= \hat{D}-i\hbar a \hat{H}\,,\qquad 
e^{b\hat{K}}\hat{D}
e^{-b\hat{K}}= \hat{D}+i\hbar b \hat{K}\,,&\\&
e^{a\hat{H}}\hat{K}
e^{-a\hat{H}}=\hat{K}-2ia\hbar\hat{D}-a^2\hbar^2\hat{H}
\,,\qquad
e^{b\hat{K}}\hat{H}
e^{-b\hat{K}}=\hat{H}+2ib\hbar\hat{D}-b^2\hbar^2\hat{K} \,,&\\&
\label{UnitaryD}
e^{c\hat{D}} \hat{H} e^{-c\hat{D}}=e^{ic\hbar}\hat{H}\,,\qquad
e^{c\hat{D}} \hat{K} e^{-c\hat{D}}=e^{-ic\hbar}\hat{K} \,,
&
\end{eqnarray} 
where the constant real parameters $a$,  $b$ and $c$ 
 have the dimensions 
 $M^{-1}L^{-1}T^{2}$, $M^{-1}L^{-2}$ and  $M^{-1}L^{-1}T$, respectively, 
 in terms of units of mass, $M$, length, $L$, and time, $T$.
 Using 
  them 
 one finds 
 \begin{eqnarray}
 \label{CBTApp}
 \hat{\mathfrak{S}}_\Omega(2i\Omega \hat{D})
\hat{\mathfrak{S}}_\Omega^{-1}=\hat{H}+\Omega^2\hat{K}\,,\qquad  
 \hat{\mathfrak{S}}_\Omega=e^{-\frac{\Omega}{\hbar}\hat{K}}e^{\frac{\hat{H}}{2\hbar\Omega}}
 e^{\frac{i}{\hbar}\ln(2)\cdot \hat{D}}\,,
\qquad\Omega\in\R^{+}\,,
 \end{eqnarray} 
 that is  just an equivalent way to write the CBT. 
 
Now, 
the generic Hamiltonian of the Swanson model   (\ref{SwansonInSO1}) 
can be transformed into $\hat{H}+\Omega_{\alpha,\beta}^2\hat{K}$
with $\Omega_{\alpha,\beta}^2=(\omega^2-4\alpha\beta)>0$ by employing 
the additional similarity transformation, whose form depends on the values of the parameters 
$\alpha$ and $\beta$. 
For this we first consider  separately the cases 
1) $\alpha+\beta \in \mathscr{D}_1= (-\infty,\omega)$  
and 
2) $\alpha+\beta \in \mathscr{D}_2=(-\omega,\infty)$, for which we  
introduce the  operators 
\begin{eqnarray}
\hat{T}_1=e^{\frac{(\beta-\alpha)}{\hbar}\hat{K}}e^{\frac{i}{\hbar}\ln(1-\frac{(\alpha+\beta)}{\omega})\hat{D}}\,,\qquad
\hat{T}_2=e^{\frac{i}{\hbar}\ln(1-\frac{4\alpha\beta}{\omega^2})\hat{D}}
e^{\frac{\alpha-\beta}{\hbar}\hat{H}}
e^{-\frac{i}{\hbar}\ln(1+\frac{(\alpha+\beta)}{\omega})\hat{D}}\,. 
\end{eqnarray}
In both  cases, the Swanson Hamiltonian 
is transformed into $\hat{H}+\Omega_{\alpha,\beta}^2\hat{K}$
with 
corresponding values of the parameters $\alpha$ and $\beta$,
$\hat{T}_a \hat{H}_{\alpha,\beta}\hat{T}_a^{-1}=\hat{H}+\Omega_{\alpha,\beta}^2\hat{K}$, $a=1,2$. 
It is notable that both these 
 transformations are well 
defined when $\alpha+\beta \in \mathscr{D}_1\cap\mathscr{D}_2= (-\omega,\omega)$, and they produce the same operator 
while acting on $2i\Omega_{\alpha,\beta}\hat{D}$ for these values of the parameters. 
However, by taking the  free particle realization of the conformal generators (\ref{so21gen}), 
and considering the corresponding mappings of operators  $\hat{x}$ and $\hat{p}$,    
\begin{eqnarray}
&
\hat{T}_1\hat{x}\hat{T}_1^{-1}=\sqrt{1-\omega^{-1}(\alpha+\beta)}\hat{x}
\,,\qquad
\hat{T}_1\hat{p}\hat{T}_1^{-1}=\frac{\hat{p}+im (\beta-\alpha)\hat{x}}{\sqrt{1-\omega^{-1}(\alpha+\beta)}}\,,
&\\&
\hat{T}_2\hat{x}\hat{T}_2^{-1}=(\frac{\omega^2-4\alpha\beta}{\omega^2+\omega(\alpha+\beta)})^{\frac{1}{2}}
\left(\hat{x}-\frac{i}{m}\frac{\alpha-\beta}{\omega^2-4\alpha\beta}\hat{p}\right)\,,
\qquad
\hat{T}_2\hat{p}\hat{T}_2^{-1}=(\frac{\omega^2-4\alpha\beta}{\omega^2+\omega(\alpha+\beta)})^{-\frac{1}{2}}\hat{p}\,,
&
\end{eqnarray}
  one notes that these two transformations are not equivalent. 
  Then, this is a particular  example when the Dyson map, and so the metric 
  operator, is not unique,
  compare with Ref.  \cite{Sm2006}. 
    
By combining these operators and the CBT one concludes that 
\begin{eqnarray}
&
\hat{T}_a^{-1}\hat{\mathfrak{S}}_{\Omega_{\alpha,\beta}}(2i\Omega_{\alpha,\beta} \hat{D} )\hat{\mathfrak{S}}
_{\Omega_{\alpha,\beta}}^{-1}\hat{T}_a=\hat{H}_{\alpha,\beta,\omega}\,
&
\end{eqnarray}
is
 the similarity transformation which  relates our $\mathcal{PT}$-symmetric system 
 with the Swanson model 
in the corresponding range of the parameters.

All the described similarity transformations can be extended to the supersymmetric cases 
we studied 
since the extra terms in the supersymmetric 
versions of $\hat{H}_\mathcal{PT}$ are conformal invariant.

\section{CBT and  $\phi_{n_1,n_2}(x_1,x_2)$ functions}
\label{AppPropPhi}
Functions $\phi_{n_1,n_2}(x_1,x_2)$ given by Eq. (\ref{phin1n2}) are mapped by the CBT into 
\begin{eqnarray}
&\label{LadderLandau}
\psi_{n_1,n_2}(x_1,x_2)=\sqrt{\frac{m\omega}{\hbar\pi n_1!n_2!}} H_{n_1,n_2}
(\sqrt{\frac{m\omega}{\hbar}}x_1,\sqrt{\frac{m\omega}{\hbar}}x_2)e^{-\frac{m\omega}{2\hbar}(x_1^2+x_2^2)}\,,&\\&
 H_{n_1,n_2}(\eta_1,\eta_2)=2^{n_1+n_2}\sum_{k,l=1}^{n_1,n_2}(i)^{n_1+n_2-l-k}H_{l+k}(\eta_1)H_{n_1+n_2-l-k}(\eta_2)\,.
&
\end{eqnarray}
The latter are the eigenstates of the operator $\hat{H}_L$  satisfying the orthonormality relation
$\bra{\psi_{n_1,n_2}}\ket{\psi_{n'_1,n'_2}}=\delta_{n_1n'_1}\delta_{n_2n'_2}$ \cite{ERIHO}. 
Application of  the CBT operator
$\hat{\mathscr{S}}$ defined in (\ref{CBT2x2twodim})
 to   spinors (\ref{PTLandauState})  yields  
\begin{eqnarray}
\label{eigen2dL}
\Psi_{n_1,n_2}^{(+)}=\left(\begin{array}{c}
\psi_{n_1,n_2}\\
0
\end{array}\right)\,,\qquad 
\Psi_{n_1,n_2}^{(-)}=\left(\begin{array}{c}
0\\
\psi_{n_1,n_2}
\end{array}\right)\,,
\end{eqnarray}
which are the eigenstates of $\hat{\mathcal{H}}_L$ with the eigenvalues $\hbar\omega(2n_2+1\mp 1)$. 

The basic operators $\hat{x}_\pm$ and $\hat{p}_\pm$ 
act as ladder operators on functions 
$\phi_{n_1,n_2}(x_1,x_2)$, 
\begin{eqnarray}
&\label{LadderLike1}
\sqrt{\frac{m\omega}{2\hbar}}\hat{x}_+ \phi_{n_1,n_2}=\sqrt{n_1+1}\phi_{n_1+1,n_2}\,,\qquad 
\frac{i}{\sqrt{2m\omega\hbar}}\hat{p}_- \phi_{n_1,n_2}=\sqrt{n_1}\phi_{n_1-1,n_2}\,,
&\\&\label{LadderLike2}
\sqrt{\frac{m\omega}{2\hbar}}\hat{x}_- \phi_{n_1,n_2}=\sqrt{n_2+1}\phi_{n_1,n_2+1}\,,\qquad 
\frac{i}{\sqrt{2m\omega\hbar}}\hat{p}_+ \phi_{n_1,n_2}=\sqrt{n_2}\phi_{n_1,n_2-1}\,.
&
\end{eqnarray}
Application of the CBT to these equalities   produces the relations
\begin{eqnarray}
&\label{Ladderb1}
\hat{b}_1^+ \psi_{n_1,n_2}=\sqrt{n_1+1}\psi_{n_1+1,n_2}\,,\qquad 
\hat{b}_2^-\psi_{n_1,n_2}=\sqrt{n_1}\psi_{n_1-1,n_2}\,,
&\\&\label{Ladderb2}
\hat{b}_2^+ \psi_{n_1,n_2}=\sqrt{n_2+1}\psi_{n_1,n_2+1}\,,\qquad 
\hat{b}_1^-\psi_{n_1,n_2}=\sqrt{n_2}\psi_{n_1,n_2-1}\,.
&
\end{eqnarray}
From here one also notes that functions  $\psi_{n_1,n_2}$ are the eigenstates 
of the ERIHO Hamiltonian 
$\hat{H}_\gamma$ 
with eigenvalues  $\hbar\omega[(1+\gamma)n_1+(1-\gamma)n_2+1]$. 

Equations (\ref{LadderLike1}) and (\ref{LadderLike2}) are the basic formulas that allow to compute the action of any
higher order operator appearing in Sec. \ref{SecSUSYCBT2d}.

 \section{
 Operators $\hat{p}_\pm^{(\alpha)}$  }
\label{AppPonPhiPsi}
In this Appendix we explore the possibility 
to have well-defined momenta operators for the systems with the Aharonov-Bohm flux
by 
looking for their  action on
physical eigenstates. 
First we compute
\begin{eqnarray}
\label{PonpsiJ}
\begin{array}{lcl}
\hat{p}_\pm^{(\alpha)}\phi_{\kappa,\ell}^{(\alpha)}&=&-i\hbar \kappa \sqrt{\frac{\kappa}{2\pi}}
[
\frac{\partial}{\partial\zeta} J_{|\ell+\alpha|}(\zeta)\mp \frac{\ell+\alpha}{\zeta}J_{|\ell+\alpha|}(\zeta)
] e^{i(\ell\pm 1)\varphi}\\
&=&-\frac{i\hbar \kappa}{2} \sqrt{\frac{\kappa}{2\pi}}
\left[\left(1\mp \frac{\ell+\alpha}{|\ell+\alpha|}\right)
J_{|\ell+\alpha|-1}(\zeta)-
\left(1\pm \frac{\ell+\alpha}{|\ell+\alpha|}\right)
J_{|\ell+\alpha|+1}(\zeta)
\right]
e^{i(\ell\pm 1)\varphi}\,,
\end{array}
\end{eqnarray}
where $\zeta=\kappa\rho$, 
and the second line is obtained by using the recurrence relations 
$\frac{2\beta}{\eta}J_\beta(\eta)=J_{\beta-1}(\eta)+J_{\beta+1}(\eta)$ and 
$2\frac{d}{d\eta}J_\beta(\eta)=J_{\beta-1}(\eta)-J_{\beta+1}(\eta)$.
For any arbitrary value of $\alpha$,  we choose two angular momentum quantum numbers 
$\ell_1$ and $\ell_2$ such that    
 $\ell_1+\alpha>0$, $\ell_2+\alpha<0$ and $|\ell_i+\alpha|<1$. 
 For these both cases one has  
\begin{eqnarray}
&\label{badstate}
\hat{p}_-^{(\alpha)}\phi_{\kappa,\ell_1}^{(\alpha)}=
-i\sqrt{\frac{\hbar^2\kappa^3}{2\pi}}J_{|\ell_1+\alpha|-1}(\zeta)e^{i(\ell_1-1)\varphi}\,,
\quad 
\hat{p}_+^{(\alpha)}\phi_{\kappa,\ell_2}^{(\alpha)}=
-i \sqrt{\frac{\hbar^2\kappa^3}{2\pi}}J_{|\ell_2+\alpha|-1}(\zeta)e^{i(\ell_2+1)\varphi}\,.\qquad
&
\end{eqnarray}
Since the index in the Bessel functions here are negative, one has that the obtained functions in 
(\ref{badstate}) can be  physical states if and only if $\alpha$ is an integer number. 
This follows from
 the relation $J_{-n}(\zeta)=(-1)^{n}J_{n}(\zeta)$, which is only available 
 for $n \in \Z$. As a consequence, $\hat{p}_\pm^{(\alpha)}$ are physical operators 
 for system $\hat{H}^{(\alpha)}$ only 
for integer $\alpha$.  

Now we will test the possibility to implement these 
operators for the $\mathcal{PT}$-symmetric 
system (\ref{2dPt1AB}).
In the case $\alpha>0$ one identifies two classes of states,
\begin{eqnarray}
&
\phi_{n_\rho,l}^{I(\alpha)}=\mathcal{N}_{n_\rho,l}^{(\alpha)}\rho^{2n_\rho+l+\alpha}e^{ i l\varphi}\,,\qquad l=-[\alpha],-[\alpha]+1,\ldots\,, &\\& 
\phi_{n_\rho,j}^{II(\alpha)}=\mathcal{N}_{n_\rho,j}^{(\alpha)}\rho^{2n_\rho+\alpha-j}e^{ i j\varphi}\,,
\qquad j=-[\alpha]-1,\ldots \,,
&
\end{eqnarray} 
where $[\alpha]$ is the integer part of $\alpha$.
The action of $\hat{p}_-^{(\alpha)}$ on the state $\phi_{n_\rho,-[\alpha]}^{I(\alpha)}$ yields 
\begin{eqnarray}
&
\hat{p}_-^{(\alpha)}\phi_{n_\rho,-[\alpha]}^{I(\alpha)}=-2i\hbar
\mathcal{N}_{n_\rho,-[\alpha]}^{(\alpha)}(n_\rho+\alpha-[\alpha])\rho^{2n_\rho+\alpha-[\alpha]-1}e^{-([\alpha]+1)\varphi}\,.
&
\end{eqnarray}
In the case in which $\alpha=[\alpha]$, the expression on
 the right hand side of this last equation becomes 
$\phi_{n_\rho,-([\alpha]+1)}^{II(\alpha)}$, and we have a 
 transition between these two classes of 
eigenstates. For non-integer $\alpha$,
however, 
the obtained functions do not correspond to what we understood as 
physical states\,: 
 in the case $n_\rho=0$ they are
  singular at the origin. 
As a consequence, the action of conformal bridge transformation 
on such states
will produce 
$\sim
 \rho^{\alpha-([\alpha]+1)}e^{-\frac{m\omega}{2\hbar}\rho^2}e^{-(i[\alpha]+1)\varphi}$, 
that 
 does not coincide with  
 the form of physical eigenstates  (\ref{LLLStates}) with $n_\rho=0$ and
  $\ell=-[\alpha]-1$  of the Landau problem 
 in the presence of the Aharonov-Bohm  flux. 
 
For
$\alpha=-\beta$, $\beta>0$ (that includes in this way the anyonic case)  
we also divide the eigenstates in two classes:  
\begin{eqnarray}
&
\phi_{n_\rho,l}^{I(-\beta)}=\mathcal{N}_{n_\rho,l}^{(-\beta)}\rho^{2n_\rho+l-\beta}e^{ i l\varphi}\,,\qquad l=[\beta],[\beta]+1,\ldots\,, &\\& 
\phi_{n_\rho,j}^{II(-\beta)}=\mathcal{N}_{n_\rho,j}^{(-\beta)}\rho^{2n_\rho+\beta-j }e^{ i j\varphi}\,,
\qquad j=[\beta]-1,\ldots \,,
&
\end{eqnarray} 
and  the action  of $\hat{p}_-^{(-\beta)}$
on  $\phi_{n_\rho,l}^{I(-\beta)}$
leads to a conclusion similar to that in
 the case $ \alpha> 0 $.

So,
one concludes that $\hat{p}_\pm^{(\alpha)}$
are physical operators only for 
$\alpha\in\Z$.

\end{document}